\documentclass[aps,prc,nofootinbib,amsmath,amssymb]{revtex4}

\usepackage{graphicx}
\usepackage{xcolor}

\begin{document}
\title{ Three-body model for  $K(1460)$ resonance
}
\author{I. Filikhin$^{1}$, R. Ya. Kezerashvili$^{2,3}$,  V. M. Suslov$^{1}$, Sh. M. Tsiklauri$^{4}$ and B. Vlahovic$^{1}$}

\affiliation{\mbox{$^{1}$ Mathematics and Physics Department, North Carolina Central University, }\\
Durham, NC 27707, USA \\
\mbox{$^{2}$Physics Department, New York City College
of Technology, The City University of New York,} \\
Brooklyn, NY 11201, USA \\
\mbox{$^{3}$The Graduate School and University Center, The
City University of New York,} \\
New York, NY 10016, USA\\
\mbox{$^{4}$Borough of Manhattan Community College, The City University of New York}
}
\date{\today}
\begin{abstract}
\noindent
We develop the three-body
$KK\bar K$ model for the $K(1460)$ resonance based on the Faddeev equations in configuration space.
A single-channel approach is utilized that takes into account the difference of masses
of neutral and charged kaons. It is demonstrated that the mass splitting of the $K(1460)$ resonance takes a place
around 1460 MeV according to $K^0K^0{\bar K}^0$, $K^0K^+K^-$ and $K^+K^0{\bar K}^0$, $ K^+K^+K^-$ neutral and charged particle configurations, respectively. The calculations are performed with two sets of $KK$ and $K\bar K$ phenomenological potentials, where strengths interactions are  considered the same for the isospin singlet and triplet states.
We study the effect of repulsion of the $KK$ interaction on the mass of the $KK\bar K$ system and evaluate the effect of the mass polarization.
The Coulomb interaction for description of the $K(1460)$ resonance is considered for the first time. The mass splitting in the $K$(1460) resonances is evaluated to be in range of 10 MeV with taking into account the Coulomb force.
The three-body model with the $K\bar K$ potential, which has the different strength of the isospin singlet and triplet interactions and related to the condition of obtaining a quasi-bound three-body state is also considered. Our results are in reasonable agreement
with the experimental mass of the $K(1460)$ resonance.

\end{abstract}
%
%
\maketitle
\section{\label{sec:intro} Introduction}

\bigskip Since the early 1960s, when the quark model was developed, it became
clear that hadrons are not elementary particles but composed of quarks and
antiquarks. In the classical quark model, a baryon is composed of three
quarks, and a meson is composed of one quark and one antiquark. Today the
internal structure of hadrons is a prominent topic of high energy physics
\cite{AliProg2017, Chen2017, Olsen, Mei2018}. Quarks and gluons are confined
within the mesons and baryons. Thus, hadrons are composite objects of quarks
and gluons governed by quantum chromodynamics (QCD), which has been
established as the theory describing the strong interaction. However, QCDs
application to low-energy hadron phenomenology is still relatively
unexplored, and there are open problems to be studied. The interpretation of
hadronic states is one of the most critical issues in hadronic physics,
particularly for the exotic states which cannot be easily collected as
quark-antiquarks or three quark states. In particular, some specific
resonances cannot be simply explained by the quark model and may be of a more
complex structure. Common features for descriptions of such specific
resonances are predictions for the existence of hadrons with substructures, which
are more complex than the standard quark-antiquark mesons and the
three-quark baryons of the original quark model that provides a concise
description of most of the low-mass hadrons \cite{Olsen}.

In the low energy region, where perturbative QCD does not work,
non-perturbative methods such as the QCD sum rule \cite{Nielsen}, lattice QCD
\cite{Nakahara, Bazavov}, chiral perturbation theory\ \cite{Gasser, 1, 2,3}%
, field correlator method (FCM)
in QCD \cite{Dosch, Simonov, Simonov2, 11} and so on, are needed. The list of just aforementioned  non-perturbative methods is not meant to be complete. In Ref. \cite{Bashir} survey of contemporary studies of hadrons and strongly interacting quarks using QCD's Dyson-Schwinger equations, in particular, Faddeev-type equations are employed for baryon calculations. This review complements and extends earlier reviews \cite{Roberts1, Roberts2, Roberts3, Roberts4}. One should mention the functional renormalization group approach \cite{Cyrol1} for quantitative first-principle studies of the QCD phase diagram and the hadron spectrum. Refs. \cite{Cyrol2, Cyrol3} are built the foundation for the work \cite{Cyrol1}, which constitutes a crucial prerequisite for future quantitative first-principle studies at finite temperature and finite chemical potential. The comprehensive review of the spectrum and electromagnetic properties of baryons, are described as relativistic three-quark bound states within QCD, and the review of nonperturbative light-front Hamiltonian methods are presented in Refs. \cite{Eichmann, Hiller}. We cite these works, but the recent literature
on the subject is not limited by them.

The physics of three-body systems has received attention for decades. The
general approach for solutions of the three-body problem at low energies is
based on the use of methods for studying the dynamics of three
particles in discrete and continuum spectra. Currentl,y among the most
powerful approaches are the method of Faddeev equations in momentum or
configuration spaces, the method of hyperspherical harmonics (HH), the
variational method in the harmonic-oscillator basis and the variational
method complemented with the use of explicitly correlated Gaussian basis
functions. To investigate the three-body systems in hadron physics, one
should solve the Faddeev equations \cite{Fad,Fad1}. The method of hyperspherical harmonics in
configuration \cite{Delves2, Smith, Delves3, SimonovHH, SimonovHH2, JKez1980, Kez1983, JKez1985, Roskez1993, Kez1994, Kez2001, Koslovsky2012, Kievsky2018} ( see also references herein) or momentum \cite{Jkrup1977, JKez1985, Koslovsky2012} spaces is another method
that is intensively used in a few-body physics, which despite its conceptual simplicity, offers great
flexibility, high accuracy, and can be used to study diverse quantum
systems, ranging from small atoms and molecules to light nuclei and hadrons. It
is a very challenging task to solve the Faddeev equations exactly and we are
usually introducing some reasonable approximations of the Faddeev equations,
such as the use of separable potentials, energy-independent kernels,
on-shell two-body scattering amplitudes, the fixed-center approximation
(FCA), the Faddeev-type Alt-Grassberger-Sandhas equations' \cite{AGS}. On the other hand an
application of the HH for the solution of three-body problems always relies on the reasonable convergence of
the method.

Using the unitary extensions of chiral
perturbation theory \cite{1} that is a good representation of QCD at low
energies \cite{2,3}, dynamically generated three-body resonances formed via
the meson-meson and meson-baryon interactions were intensively studied using
FCA for the Faddeev equations in Refs. \cite{4,41,17,18,5,6,7,Bayar,Bayar2,8,9,91,92,922,93}. Indeed, the use of chiral dynamics
in the Faddeev equations within a fixed-center approximation allows the
description of three-body resonances consisting of mesons and baryons.
Thus, hadronic composite states are introduced as few-body systems in hadron
physics \cite{12}. Therefore, interpretation of the states found in the
system of mesons and baryons became one of the challenges in theoretical
physics \cite{13}. 

The FCM is a promising formulation of the non-perturbative QCD that gives
additional support to the quark model assumptions. Progress was made \cite%
{101,102,103,10} towards placing the computation of baryon masses within the
FCM by describing a three particle system within the HH method.
Using the HH approach the ground and $p$-wave excited
states of $nnn$, $nns$, and $ssn$ baryons can be obtained \cite{10, 10P} in the
framework of the FCM \cite{11}.

It is interesting to consider a dynamical generation of $K$(1460)
pseudoscalar resonance formed by a system of three kaons. In particular,
noteworthy is the possibility of the formation of the quasi-bound states of
three kaons. The observation of pseudoscalar resonances is of fundamental
importance towards the understanding of the meson spectrum. Let us go over a
short history of $K$(1460) pseudoscalar resonance. $K$(1460) pseudoscalar
was a subject of interest already several decades ago. The first evidence
for a strangeness-one pseudoscalar meson with a mass of $\sim $ 1400 MeV and a
width of $\sim $ 250 MeV was reported via $J^{P}=0^{-}$ partial-wave
analysis of the $K\pi \pi $ system in the reaction $K^{\pm }p\rightarrow
K^{\pm }\pi ^{+}\pi ^{-}p$ \cite{Brandenburg}. The study of this
process was carried out at SLAC, using a 13 GeV incident $K^{\pm }$ beam.
A few years later, the diffractive process $K^{-}p\rightarrow K^{-}\pi ^{+}\pi ^{-}p$ at
63 GeV was studied by ACCORD collaboration \cite{DAUM1981} and the existence of a broad $0^{-}$
resonance with a mass $\sim $ 1460 MeV may now be taken as established. However, still the PDG does not yet list it as an "established particle" \cite%
{PDG2018}. In the most recent study \cite{LHCb} intermediate decays of the $%
K(1460)$ meson are found to be roughly consistent with previous studies \cite%
{Brandenburg, DAUM1981}, with approximately equal partial widths to $\bar K%
^{\ast }(892)\pi ^{-}$ and $[\pi ^{+}\pi ^{-}]^{L=0}K^{-}$,
and its resonant nature is confirmed using a model-independent partial-wave
analysis. This resonance can be considered as a $2^{1}S_{0}$ excitation of
the kaon in a unified quark model, which leads to the mass
1450 MeV \cite{Izgur85}.

By assuming isospin symmetry in the effective kaon-kaon interactions that is
attractive for $K\bar{K}$ pair and repulsive for $KK$ pair,
the $K$(1460) pseudoscalar resonance
can be the $KK\bar{K}$ system. With this idea in mind, in Ref. \cite%
{MartJidoKanada} was performed the study of the $KK\bar{K}$ system using the
single-channel variational approach in the framework of the model
\cite{KanadaJido, JidoKanada} from one hand, and within the Faddeev
equations formalism in momentum representation. In the latter case, the two-body
on-shell $t$ matrices which describe $KK$ and $K\bar{K}$ interactions by
using the Bethe-Salpeter equation in a couple-channel approach are determined and the
on-shell factorization method, from another hand. Dynamical generation of
pseudoscalar $K$(1460) resonance was considered in Ref. \cite{Albaladejo},
by studying interactions between the $f_{0}(980)$ and $a_{0}(980)$ scalar
resonances and the lightest pseudoscalar mesons. In Ref. \cite{RKSh.Ts2014} using the single-channel
description of the $KK\bar{K}$ system in the framework of the HH method
the mass of $K$(1460) resonance was calculated.
Recently, in Ref. \cite{SHY19}, the $KK\bar{K}$ system was considered
based on the coupled-channel complex-scaling method by introducing three channels $KK\bar{K}$, $\pi \pi K$ and $\pi \eta K$.
The resonance energy and width were determined using
two-body
potentials that fit two-body scattering properties.
The model potentials having the form of one-range Gaussians were proposed based
on the experimental information about $a_{0}$ and $f_{0}$ resonances. In this
model, the $K\bar{K}$ interaction depends on the
pair isospin. In particular, the isospin triplet $K\bar{K}(I=1)$ interaction
is essentially weaker than the isospin singlet $K\bar{K}-K\bar{K}$
interaction in the channel $\pi K-K\bar{K}(I=0)$.

The aim of this paper is to systematically investigate
the $KK\bar{K}$ system in the framework of a new approach for the kaonic physics $-$ the  Faddeev equations in configuration space. This approach is suitable to formulate and solve the $KK\bar{K}$ bound state problem with the effective phenomenological  $KK$ and $K\bar{K}$ potentials \cite{KanadaJido,JidoKanada,MartJidoKanada} constructed in configuration space. Existing calculations in the momentum space are contradictory to each other. We hope to bring more clarity in study of the $KK\bar{K}$ system using the well-established method, the Faddeev equations in configuration space.  We are suggesting to consider bosonic $KK\bar{K}$ system in terms of the Efimov's physics and show the connection of the Efimov's attraction with the exchange term in the Faddeev equations. As the option of the consideration of the universality in three-body systems we suggest to use the mass polarization.
In our work we try to answer the following
questions: (i) What $KK\bar{K}$ are deeply bound? (ii) Is there any strange structure peculiar to $KK\bar{K}$ system?
We present our study of the $K$(1460) resonance in the
framework of a single-channel non-relativistic potential model using the
Faddeev equations in configuration space and considering this resonance by means of a
three-body kaonic system $KK\bar{K}$. Such consideration
allows to use $KK$ and $K\bar{K}$
potentials for description of the $KK\bar{K}$ system. In our approach these potentials are only inputs along with the masses of kaons. Following Ref. \cite{MartJidoKanada} we study the $KK\bar{K}$
system using effective phenomenological potentials but taking into account the
difference in masses of $K$ and $\bar{K}$ kaons. The latter leads to
splitting the mass of the $K(1460)$ resonance according to the following neutral or charged
particle configurations: $K^0K^0{\bar K}^0$, $K^0K^+K^-$, $K^+K^0{\bar K}^0$, $ K^+K^+K^-$.
We consider two cases for the $%
KK\bar{K}$ system. In the first one, the strengths of the isospin singlet and triplet parts
of the $K\bar{K}$ potential are the same.
Such an approach leads to a simplified version of the Faddeev equations in configuration space for three particle systems.
The second case is complicated by isospin
dependence of the $K\bar{K}$ potential, when the strength of the isospin singlet and triplet parts
of the potential are different and related by the condition of obtaining a quasi-bound three-body state.
Results of our calculations are compared with the SLAC and ACCMOR collaboration experimental values for the mass of $K$(1460) resonance \cite{Brandenburg,DAUM1981} and the recent experimental study \cite{LHCb}.


This paper is organized as follows. In Sec. II we present our theoretical model. The Faddeev equations in configuration space are formulated and we present the particle configurations in a three-body kaonic system $KK\bar{K}$. The results of calculations for the masses of different configurations, interpretation of the results, including a comparison to the previous ones obtained within different methods, are presented in Sec. III. The concluding remarks follow in Sec. IV.

\section{\label{sec:model} Theoretical model}
\subsection{Formalism}

Configuration space methods are a valuable tool for the analysis of the three-body problem with
short-range interactions \cite{Fed1, Fed2}. Considering the $K K \bar K$ system as three interacted bosons having positions $\mathbf{r}_{1},$ $\mathbf{r}_{2}$ and $\mathbf{r}_{3}$,
once the two-body interactions for the $K\bar{K}$ and $KK$ subsystems are defined, one can determine its wave function by solving the Faddeev equations.
The bound state problem for the $K K \bar K$ system we formulate by using the Faddeev equations in configuration space \cite{FM} for bosonic $AAB$ system
with two identical particles. The total wave function of the $KK\bar{K}$ system is decomposed
into the sum of the Faddeev components $U$ and $W$ corresponding to the $(KK)%
\bar K$ and $(K{\bar K})K$ types of particles rearrangements: $\Psi
=U+(I+P)W$, where $P$ is the permutation operator for two identical
particles. For a three--body system, which includes two identical bosons,
the Faddeev equations represent the set of
two equations for the components $U$ and $W$ \cite{K1991} that reads:
\begin{equation}
\begin{array}{l}
{(H_{0}^{U}+V_{KK}-E)U=-V_{KK}(W+PW),} \\
{(H_{0}^{W}+V_{K \bar K}-E)W=-V_{K \bar K}(U+PW),}%
\end{array}
\label{eq:1}
\end{equation}
where the potentials for $KK$ and $K\bar{K}$ pairs are defined as $%
V_{KK}$ and $V_{K \bar K}$, respectively. In Eqs. (\ref{eq:1}) $H_{0}^{U}$ and $H_{0}^{W}$ are the kinetic energy operators of three particles written in the Jacobi coordinates (see Appendix \ref{sec:testing0}) corresponding to the $(KK)\bar K$ and $(K{\bar K})K$ types
of the three particles rearrangements.
The total isospin of the $K K \bar K$ system is considered to be $\frac{1}{2}$. The set of particles in the $K K \bar K$ system is defined by total isospin projections, which can be $-1/2$ or $1/2$.  The possible isospin configurations with isospin $3/2$ are not taken
into account in our calculations due to the smallness of corresponding contributions.

In general, we employ the $s$-wave isospin dependent $V_{KK}$ and $V_{K{\bar K}}$ potentials having singlet and triplet components: $V_{K K}=diag\{v^s_{KK},v^t_{KK}\}$, $V_{K\bar K}=diag\{v^s_{K{\bar K}},v^t_{K{\bar K}}\}$. One should mention that due to Bose-Einstein statistics the strength of the $s$-wave $KK$ interaction $v^s_{KK}=0$, because the isospin singlet wave function of the pair is antisymmetric. Therefore, the corresponding interaction should be suppressed. The separation of isospin variables leads to the following form of the Faddeev equations:
\begin{equation}
\label{eq:1comp}
\begin{array}{l}
(H^U_0+v^{t}_{KK}-E){\cal U}
=-v^{t}_{ KK}(-\frac12 {\cal W}^t+\frac{\sqrt{3}}2 {\cal W}^s-\frac12 p{\cal W}^t+\frac{\sqrt{3}}2 p{\cal W}^s), \\
(H^W_0+v^{s}_{K\bar{K}}-E){\cal W}^s=-v^{s}_{K\bar{K}}((\frac{\sqrt{3}}2 {\cal U}+\frac12p{\cal W}^s - \frac{\sqrt{3}}2 p{\cal W}^t), \\
(H^W_0+v^{t}_{K\bar{K}}-E){\cal W}^t
=-v^{t}_{K\bar{K}}(-\frac{1}2{\cal U}-\frac{\sqrt{3}}2 p{\cal W}^s-\frac12 p{\cal W}^t).
\end{array}
\end{equation}
The  singlet and triplet ${\cal W}$ components of the wave function are noted by indexes $s$ and $t$, respectively, $U$ is the triplet component,  and the
exchange operator $p$ acts on the particles' coordinates only.  Within $s$-wave approach, the equation for the singlet component corresponding to the $KK$ singlet potential is omitted due to the isospin symmetry. The similar property is demonstrated for equations describing $AAB$ systems like $NN\bar {K}$ \cite{K2015, Revai} and $nnp$ \cite{Igor2019}.


\begin{table}[b]
\caption{\label{tab1} Sets of parameters for the $K\bar {K}$ and $KK$
potentials. }
\centering
\begin{tabular}{ccccc}
\hline
& \multicolumn{2}{c}{A ($b$=0.47 fm)} & \multicolumn{2}{c}{B ($b$=0.66 fm)} \\ \hline
Interaction & $v^{s},$ MeV & $v^{t},$
MeV&  $v^{s},$ MeV & $v^{t},$ MeV \\ \hline
$K\bar{K}$ &$-1155-283i$ & $-1155-283i$ & $-630-210i$  & $-630-210i$
\\
$KK$  & $0$ & $313$ & $0$ & $104$ \\
 \hline
\end{tabular}
\end{table}

 For the description of the effective kaon-kaon interaction
we use the potentials from Refs. \cite{KanadaJido,JidoKanada,MartJidoKanada}
that are written in one-range Gaussian form as $V_{A}(r)=\sum\limits_{I=0,1}V_{A}^{I}\exp \left[ -\left(
r/b\right) ^{2}\right] P_{A},$ where $b$\ is
the range parameter having the same value as for the
$\bar{K}N$ interaction, $P_{A}$ the isospin projection operator, and
the index $A$ is related to the type of interaction $A\in KK$,
$K\bar{K}.$ The strength of strongly attractive $s$-wave
$K\bar{K}$ interaction in the isospin singlet and triplet states $V_{K\bar{K}}^{I=0,1}$=$v^s_{K\bar K}$=$v^t_{K\bar K}$
=$-1155-i283$ MeV with $b=0.47$ fm and $V_{K\bar{K}}^{I=0,1}$=$v^s_{K\bar K}$=$v^t_{K\bar K}$
=$630-210i$ MeV with $b=0.66$ fm are considered the same for the isospins $I=0$
and $I=1$ \cite{KanadaJido,JidoKanada,MartJidoKanada}.
Considering that $KK$ and $\bar{K}\bar{K}$
interactions are isospin invariant and that there are no open decay channels for
the $\bar{K}\bar{K}$ system, the $KK$ potential is a real. The
strength of the $s$-wave $KK$ interaction for $I=0$ is $V_{KK}^{I=0}=v^s_{KK}=0$
and for isospin $I=1$ it has a
relatively weak repulsion that is considered as $V_{KK}^{I=1}=v^t_{K\bar K}
=313$ MeV and $V_{K\bar{K}}^{I=1}=v^t_{K\bar K}=104$ MeV for parameterizations with $b=0.47$ fm and $b=0.66$ fm,
respectively. As mentioned above, the choice of the range parameters $b$ is related to the
description of the $\bar{K}N$ interaction. The value $b=0.66$ fm for
the effective $\bar{K}N$ interaction corresponds to the effective
Akaishi-Yamazaki potential derived in Refs. \cite{AYPRC2002, AYKN}
phenomenologically by using $\bar{K}N$ scattering and kaonic hydrogen
data and reproducing the $\Lambda(1405)$ resonance as the $\bar{K}p$
bound state at 1405 MeV. This potential is energy independent. The value
$b=0.47$ fm for the effective $\bar{K}N$ interaction corresponds to
the potential obtained in Ref. \cite{DHW} within the chiral
SU(3) effective field theory and is derived based on the chiral unitary approach
for the $s$-wave scattering amplitude with the strangeness $S=-1$. This potential
reproduces the total cross sections for the elastic and inelastic
$\bar{K}p$ scattering, threshold branching ratios, and the $\pi\Sigma$
mass spectrum associated with the $\Lambda$(1405). The strength of
${K}\bar{K}$ interaction was determined by fitting masses of the $f_{0}%
$(980) and $a_{0}$(980) resonances with the input width 60 MeV
\cite{JidoKanada}. The strength of the $KK$ interaction was obtained in
Ref. \cite{KanadaJido} to reproduce the $KK$ scattering length given by a
lattice QCD calculation in Ref. \cite{QCDcalculKK}. Following Ref.
\cite{JidoKanada} we refer to the kaon-kaon interaction potential with
$b=0.47$ fm and $b=0.66$ fm as A and B, correspondingly.
The set of values of the
potential strength $V_{A}^{I}$ for each interaction and two optimized values
for the range parameter (sets A and B, respectively) are given in Table \ref{tab1}.

Taking into account that the potentials have the same components in isospin singlet and triplet states, the Faddeev equations (\ref{eq:1comp}) can be reduced using an algebraic transformation \cite{K2015} defined by the diagonal matrix
to the following form:
\begin{equation}
\label{eq:1av}
\begin{array}{l}
   (H_0^U+v^t_{KK}-E){\cal U}
  =-v^t_{KK}(1+p){\widetilde{ \cal W}} , \\
   (H_0^W+v_{K\bar K}-E){\widetilde{ \cal W}}=-v_{K\bar K}
  ({ \cal U}+p{\widetilde { \cal W}}).
\end{array}
\end{equation}
The diagonal matrix $D =(-\frac{1}{2}, \frac{\sqrt{3}}{2})$  defines this transformation and
the transformation related to the $K\bar K$ potential is given as $v_{K\bar K}=DV_{K\bar K}D^T$, where $V_{K\bar K}=diag\{v^s_{K{\bar K}},v^t_{K{\bar K}}\}$.
The corresponding Faddeev components are $\widetilde{{ \cal W}}=D{\cal W}$, where ${\cal W}=({\cal W}^s,{\cal W}^t)^T$.

For the three-body system described by Eqs. (\ref{eq:1av}) one can evaluate  the mass polarization using the definition:
\begin{equation}
\label{eq:delta}
\Delta =2E_2-E_3(V_{KK}=0).
\end{equation}
Here, $E_2$ is $K\bar K$  two-body  energy  and $E_3(V_{KK}=0) $
is the three-body energy, when the ${KK}$ interaction between identical particles is omitted. The value of $\Delta$ is positive  and  depends  on the mass ratio  of the particles \cite{H2002,FG2002,F2018}.

When in the system $K K \bar K$ at least two particles are charged, the Coulomb interaction should be considered. The Coulomb potential can be included as a perturbation of the Hamiltonian in the left-hand side of Eq. (\ref{eq:1av}).  We present the structure of the set of Faddeev equations by taken into account with the Coulomb interaction in Appendix \ref{sec:testing0}.

The complete isospin model is based on Eqs. (\ref{eq:1comp}) with the splitting of the $K\bar K$ potential to two isospin channels $I$=0 and $I$=1, which have differ strength of the $K\bar {K}$ interaction.
The splitting of the singlet and triplet components proposed in Ref. \cite{SHY19} can be expressed by a ratio of strength parameters for the components of potential,
$V_{K\bar K}^{I=1}/V_{K\bar K}^{I=0}$. Eqs. (\ref{eq:1av}) describe the case, when $v^{t}_{K\bar{K}}/v^{s}_{K\bar{K}}=1$. This case corresponds to the $AAB$ system without spins and isospins (bosonic isospinless system) and the $KK\bar {K}$ demonstrates properties of such a system.

\subsection{Particle configurations in $K K \bar K$ system}
One can consider different particle configurations in the $K K \bar K$ system. The configurations are differed by sets of masses and pair potentials.
The Coulomb potential has to be included for a description of some configurations.
To select the configuration, we used the difference between the masses of kaons presented in Table. \ref{t1}.
\begin{table}[ht]
\caption{Kaons and  anti-kaons with the mass deference and isospin projections.
}
\label{t1}
\begin{center}
{\begin{tabular}{cccc}\noalign{\medskip} \hline \noalign{\medskip}
  Particle(Anti-particle) &  Quarks& Mass (MeV) & Isospin projection  \\ \hline \noalign{\medskip}
$K^+$ ($K^-$)       & us  & $493.7$ & $1/2$ (-1/2) \\   \noalign{\medskip}
$K^0$ ($\bar K^0$)       & ds  & $497.6$& -1/2 ($1/2$)\\     \hline
\end{tabular} }
\end{center}
\end{table}
These configurations are the following:
$K^0K^0{\bar K}^0$, $K^0K^+K^-$,  $K^+K^0{\bar K}^0$,  $ K^+K^+K^-$. Using the charge-isospin basis notations, the configurations can be identified as
$--+$, $-+-$, $+-+$, $++-$. Thus, the first two configurations correspond to the states with the projection of total isospin $-1/2$ of the  $K K \bar K$ system, while the last two have the  total isospin projection $+1/2$. Each configuration is represented as $AAB$ system -- the system with two identical particles and can be described by the Faddeev equations (\ref{eq:1comp}) and (\ref{eq:1av}): the case when the strength of the isospin components of the $K\bar {K}$ potential are differed and when $v^{t}_{K\bar{K}}=v^{s}_{K\bar{K}}=1$, respectively.

\section{\label{sec:results} Results}
Our interest is to examine the possibility of the existence of bound kaonic states in the $K K \bar K$ system. For this purpose, we solve
numerically differential Eqs. (\ref{eq:1comp}) in the case of the different strength for the isospin singlet and triplet components of $K\bar {K}$ potential and Eqs. (\ref{eq:1av}), when $v^{t}_{K\bar{K}}=v^{s}_{K\bar{K}}$. The differential Faddeev equations have been formulated in the pioneering work of Noyes and Fiedeldey \cite{NF1968} for the simplest case of $s$-wave three-particle scattering and have been generalized in Ref. \cite{MGL}. Our numerical procedure for the solution of the Faddeev equations in configuration space
is based on the finite difference approximation with spline collocations \cite{Su1996, Sus20}.
\begin{table}[b]
\centering
\caption{ The mass of the $K$(1460) resonance for the potentials of the A and B parameter sets (without the Coulomb force), A$_c$ and B$_c$ (the Coulomb force is included),   $M=\sum_\gamma m_\gamma-|E_3|$, $m_\gamma$ is kaon mass, $\gamma=1,2,3$. The energy of $KK\bar K$ quasi-bound state ($I=1/2$)  is $E_3$. The $E_2$ is the energy of bound $K\bar K$ pair. The masses and energies are given in  MeV. $\delta=\Delta/|E_3(V_{KK}=0)|$ is the relative contribution of the mass polarization. The result of Ref. \cite{MartJidoKanada} is given in parentheses. Upper bound for mass of  three-body quasi-bound state is shown as $\sum_\gamma m_\gamma-|E_2|$. $m_{K}$ is averaged kaon mass.}\label{tab-1}
\begin{tabular}{@{}*{6}{ccclcccccc}}
 \hline
Resonance&System& Particle &   Model &       $E_3$  &       $E_3(V_{K K}=0)$     &    $ E_2$       & Mass  &     Mass & Mass upper  \\
& $AAB$ & masses& &                &         &         & polarization $\delta$ (\%) &   $M$  & bound \\
  \hline
${K}(1460)$&$KK\bar K$ &                    496.0  \cite{MartJidoKanada}&    A   &     -19.8   (-21)      &  -32.1     &        -11.25    &      30.0          &                      1468.2 &1476.7\\
                     &                     &                         495.7       &   A   &    -19.7              &  -31.9     &              -11.18           &      29.9          &                      1467.4 &  1475.9   \\
                     &                     &                                            &   B   &    -22.2              &  -29.4    &              -11.17           &      24.0          &                      1464.9 &  1475.9   \\
\hline
${K^0}(1460)$& $K^0K^0{\bar K}^0$ &497.6 & A     &       -20.4          &     -33.0     &        -11.61   &         29.6          &                  1469.7 &1481.2\\
&                           & & B      &      -22.8            &  -30.1      &       -11.45     &       23.9          &                  1467.3&1481.3\\
\hline
${K^0}(1460)$& $K^0K^+K^-$&$m_{K^-}$=493.7, &  A   &          -19.3  &        -31.3     &        -10.96   &       29.7        &          1465.8  &1474.1       \\
&                           &$m_{K}$=495.7  &B                          &    -21.9    &        -29.0     &        -11.03   &       23.8           &     1463.2    &     1474.1       \\
 \hline
${K^+}(1460)$ &$K^+K^0{\bar K}^0$ & $m_{\bar K^0}$=497.6,&  A   &         -20.1   &      -32.5        &        -11.40   &       29.8        &                        1468.9 & 1477.6\\
&                           &$m_{K}$=495.7  &B                          &       -22.5  &          -29.8  &        -11.34   &       23.9          &                    1466.5& 1477.7\\
 \hline
${K^+}(1460)$&$ K^+K^+K^-$ &493.7 & A     &       -18.9  &          -30.9     &        -10.74     &     29.5             &                   1462.2 &1470.4   \\
&                         & & A$_c$      &    -20.9      &     --     &     --       &        --          &        1460.2         &-- \\
&                         & & B       &     -21.6     &       -28.7     &       -10.87      &    24.3              &                  1459.5&   1470.2\\
&                         & & B$_c$      &   -23.3    &       --     &     --       &        --          &        1457.8 &   --      \\ \hline
\end{tabular}
\end{table}

\textbf{Case when the strength of $v^{t}_{K\bar{K}}=v^{s}_{K\bar{K}}$.} For this case, results of our
calculations for the binding energy and the mass for the $KK\bar K$
system are presented in Table \ref{tab-1}. In this Table are also given the result from Ref. \cite{MartJidoKanada} obtained using the variational method for the single channel three-body potential model with the same two-body effective $KK$ and $K{\bar K}$ interactions.
The
total mass the $KK\bar K$ system is ranged from 1463.8 to 1469.4
MeV, when we consider the same $K$ meson mass $m_{K}=496$ MeV as in Refs. \cite
{MartJidoKanada, RKSh.Ts2014}. The quasi-bound state for the $KK\bar K$ with spin-parity $0^{-}$ and total isospin 1/2 is found
below the three-kaon threshold.

Studying the various particle configurations of the $KK{\bar K}$ system and their dependence on the particles masses, we are considering the following kaonic masses:
the mass $m_K$=496 MeV \cite{MartJidoKanada} corresponds to the $K^0K^0{\bar K}^0$ system, where
$m_K=(2m_{K^+}+m_{\bar K^0})/3$; the mass $m_K$=493.7 MeV corresponds to the $ K^+K^+K^-$ system;
the mass $m_K$=497.6 MeV corresponds to the $K^0K^0{\bar K}^0$ system. For the $K^0K^+K^-$ and $K^+K^0{\bar K}^0$ configurations we used the averaged mass of kaons for particle pair $K^0K^+$, $m_K$=495.7 MeV.

The difference of the masses for $K^0$ and $K^+$ violates the $AAB$ model with two identical particles of $KK\bar K$ system.  However, the approach with the averaged mass is completely satisfied with the $AAB$ model due to the set of the proposed potentials.
The Coulomb potential acting in the $K^0K^+K^-$ system also violates the $AAB$ symmetry and, therefore, we omitted the consideration of the Coulomb force for the $K^0K^+K^-$ configuration.  The Coulomb potentials in the $ K^+K^+K^-$ system
were included in the calculations  due to correspondence to the $AAB$ symmetry. A brief description of the Faddeev equations in configuration space with the Coulomb force acting in the $KK\bar K$ is given in Appendix \ref{sec:testing0}.

The comparison of our energy $E=-21$ MeV with the result obtained in Ref. \cite{MartJidoKanada}, shows some disagreement. We assume that it could be related to the numerical methods utilized to solve the corresponding equations. In Ref. \cite{MartJidoKanada} is used variational method, which depends on a choice of initial basis functions. We use direct numerical method \cite{Sus20} for the solution of the Faddeev equations in configuration space. Here, the direct solution means a solution method based on the finite-differential approximation of the boundary problem for eigenvalues with the discretization of the coordinate space. The analysis of the method is performed in Appendix \ref{sec:testing1}, where two cross-check tests are given. The first is related to the test of our computer codes for the solution of a problem similar to the one considered for the $KK\bar K$ system with the comparison with results of others authors. In the second one, we proposed an alternative way to solve the Faddeev equations by using the cluster reduction method \cite{CRM,CRM1}. The both tests evidenced that the accuracy for the results listed in Tables \ref{tab-1} and \ref{tab-3a} is reached in our calculations.

Let us continue the analysis of data from Table \ref{tab-1}, where also are presented the results of calculation for the $E_3(V_{K K}=0)$. The $E_3(V_{K K}=0)$ is defined as the energy of the quasi-bound state of the three-body system  when the repulsive
$KK$ interaction is omitted. In this case, the set of Eqs. (\ref{eq:1av}) is reduced to the single equation for the  ${\widetilde { \cal W}}$ Faddeev component:
\begin{equation}
\label{eq:W}
(H_0^W+v_{K\bar K}-E){\widetilde{ \cal W}}=-v_{K\bar K}
  p{\widetilde { \cal W}}.
\end{equation}%
The exchange term presented at the right-hand side of the equation provides the existence of a bound state with energy $E_3(V_{K K}=0)$.  In Table \ref{tab-1} are shown the two-body energy of bound pair $E_2$ and three-body energy $E_3(V_{KK}=0)$.
  Based on the analysis performed in  Ref. \cite{F2018} and according to Eq. (\ref{eq:delta}), the relation between $E_2$ and $E_3(V_{KK}=0)$ can be rewritten as
\begin{equation}
\label{eq:le2}
|E_3(V_{K K}=0)/E_2|>2 .
\end{equation}%
The results of the calculations given in Table \ref{tab-1} are in agreement with this relation. In nuclear physics, this relation is called as the "mass polarization effect" \cite{{F2018}}.
In terms of the  Efimov physics \cite{Ham2010, NE17}, the  relation (\ref{eq:delta}) is explained by the Efimov  attraction as result of a mediated attraction between  two  particles by the exchange of the  third particle.
Note that an expression, which is similar to Eq. (\ref{eq:le2}), has been previously  obtained in Ref. \cite{PP10} for bosonic two-dimensional $AAB$ systems to describe the relation of two-body and three-body energies.
Interestingly enough to note that due to universality, Efimov physics applies to virtually any field of quantum physics, be it atomic and molecular physics, nuclear physics, condensed matter or even high-energy physics (see, for example, \cite{BGelman, Ham2008}).

\begin{figure}[ht]
\includegraphics[width=19pc]{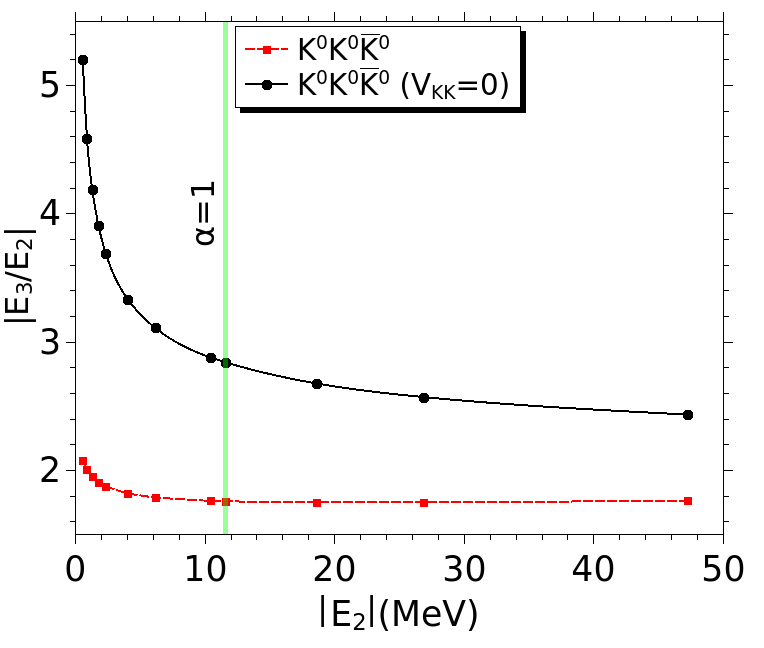}
\includegraphics[width=19pc]{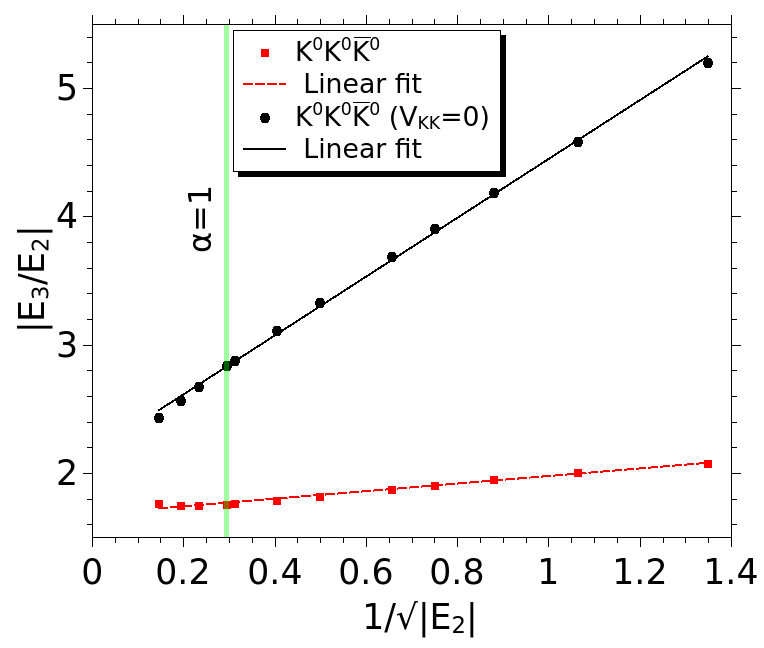}
\caption{\label{fig:01} The dependence of the ratio $E_3/E_2$ on two-body energy $|E_2|$ (left panel) and the value $1/\sqrt{|E_2|}$, which is proportional to the two-body scattering length $a_2$ (right panel) for the $K^0K^0{\bar K}^0$ system. The calculations are performed using the potentials with parameters A.
These are the parametric dependencies. The parameter is the factor $\alpha$ which defines the scaled $K\bar K$ potential as $v_{K\bar K}\to \alpha v_{K\bar K}$. The vertical line corresponds to $\alpha=1$.
}
\end{figure}

\begin{figure}[ht]
\begin{center}
\includegraphics[width=26 pc]{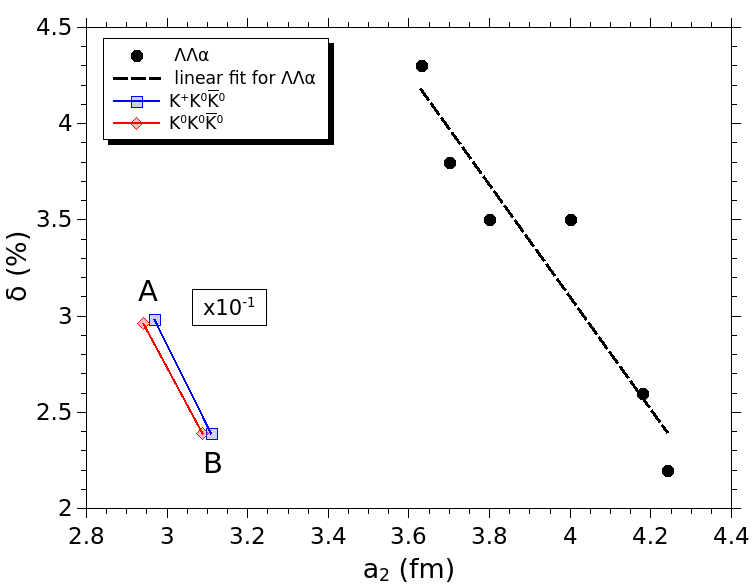}
\end{center}
\caption{\label{fig:0} The correlation between the relative contribution of the mass polarization $\delta$ and two-body scattering length $a_2$ for the $\alpha\Lambda\Lambda$ and $KK\bar K$ systems calculated with different pair potentials. The mass ratios are $m_\Lambda/m_\alpha\approx 1/4$ and $m_K/m_{\bar K}\approx 1$, respectively.
}
\end{figure}

The relation (\ref{eq:le2}) agrees with so called "Efimov scenario" \cite{NE17} defined  for  the model situation when pair potential is simply  scaled by a multiplicative factor. To illustrate this fact, in Fig. \ref{fig:01}, we present the results of calculations for the dependence of the ratio $E_3/E_2$ on two-body energy $|E_2|$ (left panel) and the value $1/\sqrt{|E_2|}$ (right panel). The $1/\sqrt{|E_2|}$ coincides with two-body scattering length $a_2$ due to approximation $|E_2|
\approx \hbar^2/(m_Ka_2^2)$. These dependencies are obtained by introducing the scaled factor $\alpha$, which parameterizes the $K\bar K$ potential and scales it as  $v_{K\bar K}\to \alpha v_{K\bar K}$. Therefore, it differs by a multiplicative factor $\alpha$. These dependencies are parametrically obtained. The parameter is the multiplicative factor of $\alpha$, defining the scaled $K\bar K$ potential.
The region of the Efimov physics corresponds to small values of $|E_2|$ (large values of $1/\sqrt{|E_2|}$). Within this region, the ratio of $E_3/E_2$ quickly increases, and a possibility for an excited state is opening.
In Fig. \ref{fig:0} we show the result for the A parameter set of the $K\bar K$ potential ($\alpha=1$). The corresponding state of the $K^0K^0{\bar K}^0$ system is far from the Efimov states. The ratio $E_3/E_2$ asymptotically approaches 2. The repulsive $KK$ potential makes the $E_3/E_2$ ratio to be smaller than the $E_3(V_{KK}=0)/E_2$ ratio. The strength of the  $KK$ repulsion defines the difference. What will happen if the $KK$ interaction would be attractive? It is clear that an attractive  $AA$ potential will make the ratio $E_3/E_2$ larger than $E_3(V_{AA}=0)/E_2$. An example of such a situation with an attractive $AA$ potential is given in Appendix \ref{sec:testing1}.

Following Ref. \cite{F2018} we evaluated the relative contribution of the mass polarization $\delta=\Delta/|E_3(V_{KK}=0)|$ to the energy of the $KK\bar K$ quasi-bound state for different physical particle configurations. The corresponding results are presented in Table \ref{tab-1}. The value of $\delta$ depends on two factors: the mass ratio of kaons and the type of the $K\bar K$ potential. When the mass ratio approaches one, and the dependence of the mass polarization on the particle configuration is hidden for the considered systems. The second factor is more significant here. One can see the dependence by comparing the results for the potentials of the A and B  parameter sets.

Summarizing the comparison, we conclude that the
mass polarization effect for the potential of the A parameter set is about 30\% and for the set B is about 24\%. There is a correlation between of two-body scattering length $a_2$ obtained with the potential that bounds non-identical particles and
the relative contribution of the mass polarization $\delta$  \cite{F2018}. The larger scattering length corresponds to the smaller mass polarization.
For the $KK\bar K$ system, the potentials of the B parameter set demonstrate larger scattering length and smaller mass polarization. This relation is shown in Fig. \ref{fig:0} along with the results obtained in Ref. \cite{F2018} for the $\alpha\Lambda\Lambda$ system with phenomenological potentials having different scattering parameters.
The relatively small mass polarization in the $\alpha\Lambda\Lambda$ system is clarified by domination of the $\alpha$-particle mass in the system,  due to the ratio $m_\Lambda/m_\alpha<<1$.
 The correlation between of the relative contribution of the mass polarization $\delta$ and two-body scattering length is only approximately linear, because the dependence of two-body parameters on the strength of a potential is more complex than the parametric dependence of a potential on the strength parameter considered above.

To show the difference between of the A and B parameter sets for the $K\bar K$ potential, we averaged Eq.  (\ref{eq:W}): $\langle H_0^W \rangle +\langle v_{K\bar K}\rangle+\langle v_{K\bar K}p \rangle=E_3(V_{K\bar K}=0)$.
To evaluate the averaged kinetic energy $\langle H_0^W \rangle$, the method proposed in Ref. \cite{FV}  was used. We considered Eq. (\ref{eq:W}) with  scaled  kaon masses by a factor $\gamma$ within the small vicinity of the point $\gamma$=1. The energy becomes a function of the $\gamma=1\pm \Delta\gamma$ and $dE(\gamma)/d\gamma=-1/\gamma^2\langle H_0^W \rangle$. The linear approximation for this derivation gives an evaluation of the averaged kinetic energy. The exchange term $\langle v_{K\bar K}p \rangle$ depends on the mass ratio and does not contribute to $\langle H_0^W \rangle$ as one can see from the numerical results listed in Table \ref{tab-1}. Using the pattern $\langle H_0^W \rangle +\langle v_{K\bar K}\rangle+\langle v_{K\bar K}p \rangle=E_3(V_{K\bar K}=0)$, the results of averaging can be written as: $274-214-93=-33$ and  $182-142-70=-30$ for the potentials of the A and B parameter sets, respectively. Here, all values are given in MeV. We see that the potentials in  the set A are "stronger" because they act on shorter distances with the larger strength. We can assume that the $KK\bar K$ system described with the A parameter set potentials is more compact.

One can assume that the contribution of the exchange term $\langle v_{K\bar K}p \rangle$ could correspond to the value evaluated by Eq. (\ref{eq:delta}) for the mass polarization term. We rewrite the expressions presented above as $274(1-214/274-93/274)=-33$ and $182(1-142/182-70/182)=-30$ for the potentials of sets A and B, respectively. The evaluation of $|\langle v_{K\bar K}p \rangle/\langle H_0^W \rangle|$ gives 34\% and
39\% for the potentials of the set A and B, respectively. These values are in disagreement with the results of the mass polarization term in Table \ref{tab-1}. Note that the mass polarization is related to the kinetic energy operator in the Schr\"odinger equation \cite{F2018}.
By using the exchange term, one cannot directly separate this kinetic part. Thus, the $\delta$ more adequately evaluates the relative contribution of the mass polarization. At the same time, one can make sure that the relative contribution of the exchange term (Efimov attraction) increases with decreasing the strength of the potential according to the "Efimov scenario".

In Table \ref{tab-1} we also present the upper bounds for the mass of three-body quasi-bound state calculated as $\sum_\gamma m_\gamma-|E_2|$, where $m_{\gamma}$, $i$=1,2,3 are the kaons masses. The values define a maximal value for the three-body resonance mass when the quasi-bound state is approximately located on the two-body threshold. Obviously, the calculated mass $M$ of the three-body resonance is less than the mass of the upper bound state.

\begin{figure}[ht]
\includegraphics[width=22pc]{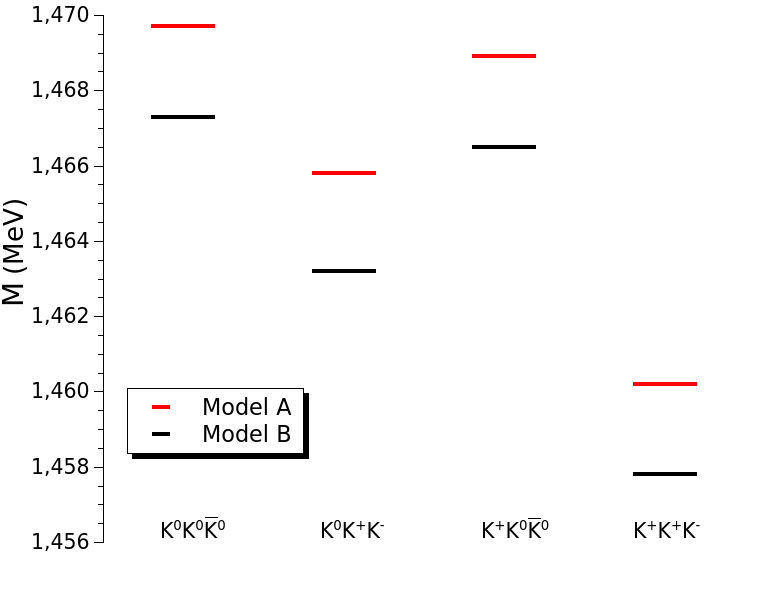}
\includegraphics[width=14pc]{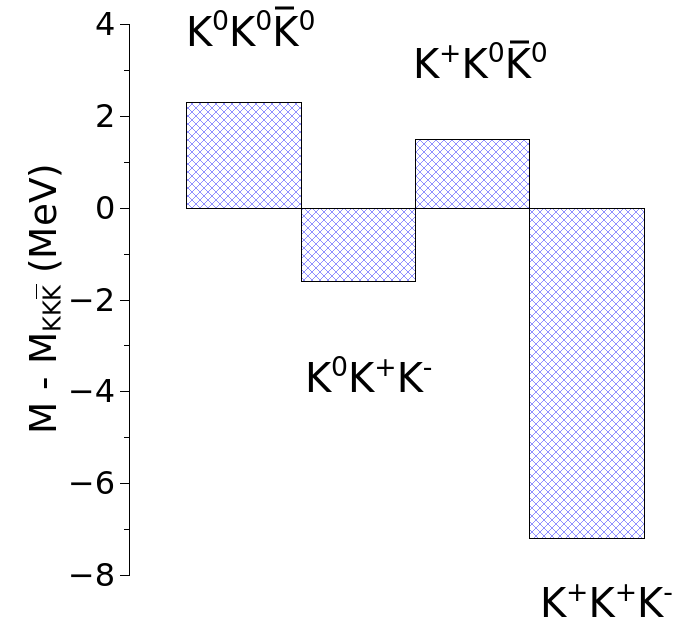}
\caption{ Mass spectrum for different particle configurations of the $KK\bar K$ system calculated with the A and B parameter sets for pair potentials (left panel). The  mass difference $M-M_{KK\bar K}$ for different particle configurations calculated with the set A for parameters of potentials (right panel).
} \label{fig:1}
\end{figure}

The mass spectrum and the mass difference for different particle configurations of the $KK\bar {K}$ system are shown in Fig. \ref{fig:1}.   The left panel presents the mass spectrum for different particle configurations of the $KK\bar K$ system calculated with the A and B parameter sets of pair potentials. It is worth to noticing that the spectrum obtained for the set of parameters B is shifted by 2 MeV relative to the spectrum obtained using the potentials with the set of parameters A. This shift is not surprising because the two-body quasi-bound energy of the model is approximately larger by 2~MeV and depends on two-body attractive $K{\bar K}$ potential. It is interesting enough to consider the difference between the average isospin model, without taking into account the difference of kaons masses and particle configuration model. The right panel in Fig. \ref{fig:1} presents the mass difference $M-M_{KK\bar K}$ for different particle configurations calculated with the parameters of pair potentials for set A. One can note that it varies from 1 to 7 MeV for different particle configurations.
\begin{figure}[ht]
\includegraphics[width=16pc]{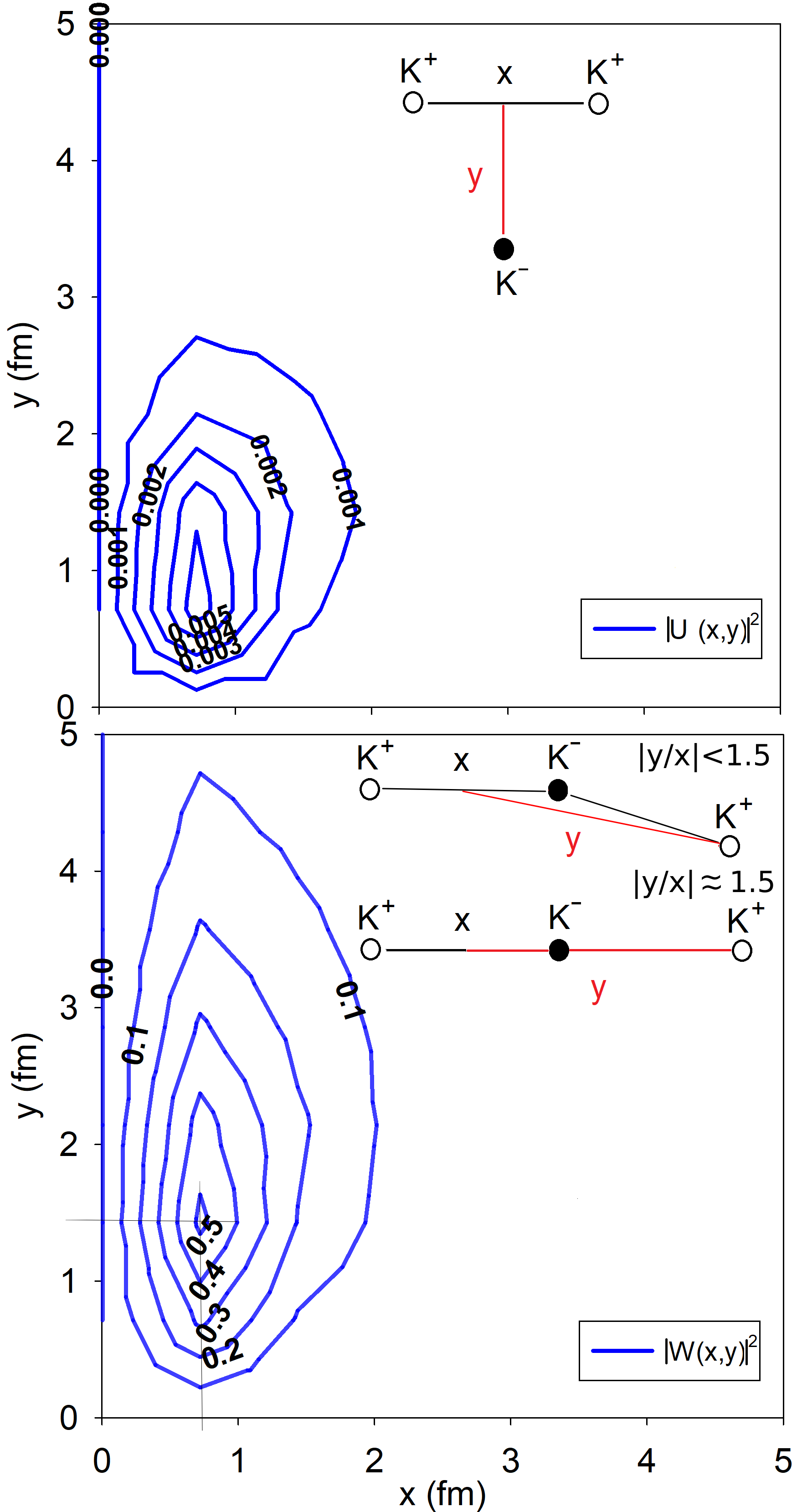}
\caption{\label{fig:FC} The probability for distribution of the particles in the $ K^+K^+K^-$ system. The contour plot of the squared modulus of the Faddeev component ${|\cal U}(x,y)|^2$  (upper panel)
and ${|\cal W}(x,y)|^2$ (lower panel) versus the corresponding Jacobi coordinates. The inserts show the most probable configurations of the particles. The $y$ coordinate is marked by the red color. For  the $U$ rearrangement ($K^{+}$+$K^{+}$)+$K^{-}$, (upper panel) the Jacobi coordinate $x$ connects of $K^{+}$ and $K^{+}$ mesons, while for the $W$ rearrangement $K^{+}$+($K^{+}$+$K^{-}$), (lower panel) the Jacobi coordinate $x$ connects $K^{+}$  and $K^{-}$ mesons.}
\end{figure}
\begin{figure}[ht]
\includegraphics[width=25pc]{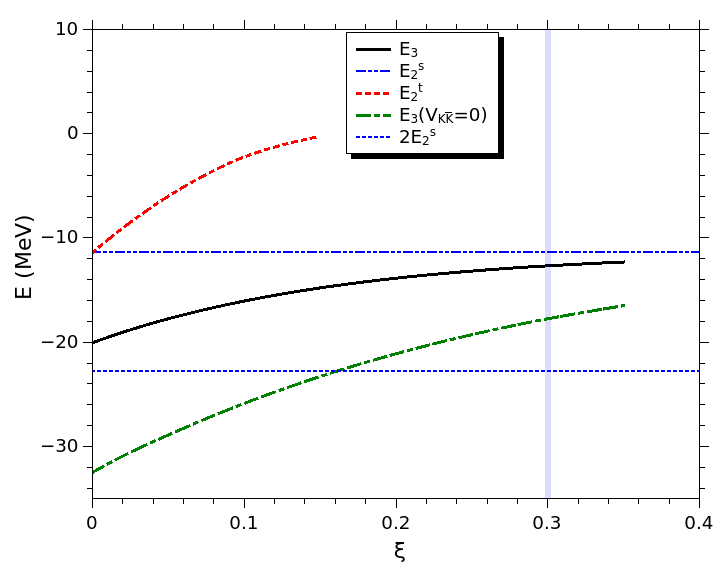}
\caption{ \label{fig:4} The energy $E_3$ of quasi-resonance in the $K^+K^0{\bar K}^0$  system (potential with the A parameter set) for different values of the scaling parameter $\xi$  defined for the triplet $K\bar K$
potential as $v^t_{K\bar K}\to(1-\xi)v^s_{K\bar K}$ \cite{SHY19}.
$E_2$ energy of subsystem $K\bar K$ (isospin singlet ($s$) or isospin triplet ($t$)). The vertical line corresponds to the value of the ratio  $v^t_{K\bar K}/v^s_{K\bar K}$ proposed in \cite{SHY19}.
}
\end{figure}

\begin{table}[ht]
\centering
\caption{\label{tab-3a} The mass and width of the  $K^0K^0{\bar K}^0$ resonance using the A and B parameter sets for the $KK$ and $K\bar K$ potentials. The results of Ref. \cite{MartJidoKanada} are shown in parentheses. The kaon masses are given in MeV.}
\begin{tabular}{@{}*{6}{ccc}}
 \hline
Potentials/Kaon mass &     $M$, MeV  &       $\Gamma$, MeV  \\  \hline
A/497.6 &  1469.7                                             &        105                            \\
A/496.0 &   1468.1 (1467)    &           104 (110)            \\
B/497.6 &    1467.3     &       117                    \\
 \hline
\end{tabular}
\end{table}

In Table \ref{tab-3a}, we present the results of our calculations for the mass and width of the  $K^0K^0{\bar K}^0$ resonance using the A and B parameter sets for the $KK$ and $K\bar K$ potentials. In this calculations we did not consider the mass difference of $K$ and ${\bar K}$ kaons. However, we considered the different value of the mass for the kaon as it is listed in Table \ref{tab-3a}. The comparison of our results with those obtained with the variational method \cite{MartJidoKanada} shows that the mass and width of $K$(1640) are close enough, 1467 MeV and 110~MeV   \cite{MartJidoKanada}, respectively. The alternative scenario is observed for the HH method \cite{RKSh.Ts2014} and the Faddeev calculations in the momentum representation \cite{MartJidoKanada}: the difference for the width is more than 50\%. In particular, for the Faddeev calculations in the momentum representation $\Gamma = 50$ MeV \cite{MartJidoKanada}, which coincides with results
obtained within the HH method \cite{RKSh.Ts2014}, where the width falls into the 41--49 MeV range for all sets of the $K\bar K$ and $KK$  interactions. The details of our method for evaluation of the width $\Gamma$ are given in  Appendix \ref{sec:width}.

Taking into account the difference of our results and results  \cite{MartJidoKanada}, we test our codes, that is presented in Appendix \ref{sec:testing1}. In the first test, we considered the system $npp $, which is described by Eqs. (\ref{eq:1av}) as a bosonic isospinless model by direct solution of the Faddeev equations in configuration space. The second test is related to the alternative approach for solution of the Faddeev equations in configuration space for the $KK\bar {K}$ system using cluster reduction method (CRM).
\begin{table}[ht]
\centering
\caption{\label{tab-3} The mass of the $K$(1460) resonance for potentials with the parameter sets A and B (without the Coulomb force) and the scaling parameter $\xi$=0.3, $v_{K\bar K}^t/v_{K\bar K}^s$=0.7 \cite{SHY19}.   The notations are the same as in Table \ref{tab-1}.  $\Gamma$ is the  width of three-body resonance. }
\begin{tabular}{@{}*{6}{ccccccccc}}
 \hline
Resonance&System&    Model &       $E_3$  &  $\Gamma$&    $E_3(V_{K K}=0)$     &    $ E^s_2$       &     Mass & Mass upper  \\
& $AAB$ & &                &    &     &          &  $M$    & bound \\
  \hline
${K^0}(1460)$& $K^0K^0{\bar K}^0$ &A     &       -12.9    &  70    &     -18.2     &        -11.61   &                       1479.9&1481.2\\
&                           & B      &      -14.7     & 78      &  -18.8      &       -11.45     &                       1478.1&1481.3\\
\hline
${K^0}(1460)$& $K^0K^+K^-$& A   &          -12.2  &        67& -17.1     &        -10.96   &                 1472.9  &1474.1       \\
&                           &B                          &    -14.0 &76    &        -18.0     &        -11.03   &           1471.1  &     1474.1       \\
 \hline
${K^+}(1460)$ &$K^+K^0{\bar K}^0$ &   A   &         -12.7  & 69 &      -17.8        &        -11.40   &                      1476.3 & 1477.6\\
&                           &B                          &       -14.4  &   77&       -18.9  &        -11.34   &                  1474.6& 1477.7\\
 \hline
${K^+}(1460)$&$ K^+K^+K^-$  & A     &       -11.9  &   69   &    -16.7     &        -10.74     &                      1469.2 &1470.4   \\
&                         & B       &     -13.8  &  76 &       -17.7     &       -10.87      &                 1467.3&   1470.2\\
\hline
\end{tabular}
\end{table}

It is interesting to consider the density distribution of particles for the  $KK\bar K$ system calculated in the framework of the Faddeev equations in configuration space. The spatial configuration of particles in the $KK\bar K$ system can be understood by plotting the spatial probability amplitudes, i.e., the squared modulus of the Faddeev components ${|\cal U}(x,y)|^2$
and ${|\cal W}(x,y)|^2$ in terms of the Jacobi coordinates $x$ and $y$. In Fig. \ref{fig:FC} we present calculations results of the probability distribution for the charged kaon resonance ${K^+}(1460)$ described as the $ K^+K^+K^-$ system by using the potentials with the parameter set B. In these figures are presented the contour plots of the squared modulus ${|\cal U}(x,y)|^2$ and ${|\cal W}(x,y)|^2$ in the frame of 5 fm$\times$5 fm, as well as the related spatial configurations. The careful examination of the contour plots shows that the squared modulus' maximal values differ by two order and, therefore, the corresponding spatial configurations probabilities.
The component ${\cal W}(x,y)$ is dominant in the total wave function. The localization of the particles in the system corresponds to the highest probability associated to the component ${\cal W}(x,y)$ for the $x$ and $y$ coordinates values. The coordinates' modules in the most favorable position are approximately related to the ratio of $|y|/|x|\sim$ 1.5. The squared modulus of the ${\cal W}(x,y)$ component displays very large asymmetry, being strongly elongated in
the $y$-direction. The spatial configuration presented in the inset reflects this ratio.
Let us note that the localization of $K^+$ mesons must be symmetrical relative to the $K^-$ due to
exchange symmetry for the identical $K^+$ mesons. The inset in the lower panel in Fig. \ref{fig:FC} presents two positions of the particles in the $KK\bar K$ system
that satisfy this condition. For the first position, the ratio $|y|/|x|$<1.5. The second position corresponds to the ratio $|y|/|x|$$\approx $1.5.
According to the spatial probability distribution for the component ${\cal W}(x,y)$ given in Fig. \ref{fig:FC}, the second position is the most probable.
Thus, the particles in the $ K^+K^+K^-$ system are distributed along one line like a chain-like spatial configuration ($K^+$)$-$($K^-$)$-$($K^+$). The distance between $ K^+K^-$ is 0.8 fm, while the distance between $ K^+K^+$ is twice larger. The latter is not surprising because the  $K^{+}K^{-}$ interaction is strong and attractive, while the interaction between the identical particles $K^{+}K^{+}$ is weak and repulsive. The other spatial configuration has a low probability and can be represented as a triangle spatial configuration with the basis side and height of 0.8 fm, respectively.

\textbf{Case when the strength of $v^{t}_{K\bar{K}} \ \textbf{and} \  v^{s}_{K\bar{K}}$ is differed.} Let's now focus on the dependence of three-body energy on the strength of isospin splitting of the $K\bar K$ potential. To consider this case, one should solve Eqs. (\ref{eq:1comp}). In Fig. \ref{fig:4} we present the calculation results for $K^+{\bar K}^0K^0$  systems for the potentials of the parameter set A. The splitting means that the isospin triplet component of the potential decreases as $v^t_{K\bar K}=(1-\xi)v^s_{K\bar K}$ under the condition, that the isospin singlet potential is not changed and provides the two body threshold $E_2^s$ about 11 MeV (see Table \ref{tab-3}). The value of 0.3 for the scaling parameter  $\xi$ corresponds to the proposed in Ref. \cite{SHY19} relation between the singlet and triplet components of the $K\bar K$ potential.
Our calculations show that the triplet bound state exists when the scale parameter $\xi$ is less than 0.15. Thus, the proposed in  \cite{SHY19} model assumes that the triplet state is not bound. The quasi-bound state of the $K^+K^0{\bar K}^0$  system has the energy $-12.4$~MeV, which is near the two-body threshold. Due to the isospin splitting of the $K\bar K$ interaction, the relation (\ref{eq:le2}) is invalidated and the opposite relation takes place: $|E_3(V_{K K}=0)/E^s_2|<2$. Also the value of
$2E_2-E^s_3(V_{KK}=0)$ becomes negative in contrast to the positive value of $\Delta$ in Eq. (\ref{eq:delta}).

\section{Summary}
\label{sec:summary}

We developed a new framework for the Faddeev calculations in configuration space for the $K$(1460) dynamically generated resonance in this work. Our three-body non-relativistic single channel model predicts a quasi-bound state for the $KK\bar K$ system with the mass around 1460 MeV. The calculations are performed using two sets of phenomenological $KK$ and $K\bar {K}$ potentials, when the strength of $K\bar K$ interaction has no difference in the singlet and triplet isospin states, and taking into account various particle configurations of the $KK{\bar K}$ system. Our study was extended to a more complicated case when the isospin singlet and triplet parts of the $K\bar {K}$
potential are different and related by the condition of obtaining a quasi-bound three-body state.

In our study, the mass difference between the kaons was taken into account to separate physical particle configurations of the $KK{\bar K}$ system: $K^0K^0{\bar K}^0$, $K^0K^+K^-$, $K^+K^0{\bar K}^0$, $ K^+K^+K^-$. These improvements enable us to investigate these kaonic configurations systematically, moreover the first, time the Coulomb interaction has been taken into account for description of the charged configurations. The mass splitting in the $K$(1460) resonances is evaluated to be in range of 10 MeV with taking into account the Coulomb force in the case of charged resonances. It is worth mentioning that a hypothetical chain-like spatial configuration ($K^+$)$-$($K^-$)$-$($K^+$) would constitute a favorable  structure of the $KK\bar {K}$ system.

We considered the mass polarization effect in the $KK\bar K$ system and evaluated the effect of the repulsion strength of $KK$ potential.
The mass polarization term, which is well separated in the Schr\"odinger equation as a part of the kinetic energy operator and the exchange term defined by the Faddeev equations, is evaluated and discussed. This term is closely related to the "Efimov attraction". Our calculations demonstrate that a peculiar feature of the mechanism for binding of the $KK{\bar K}$ system is the s-wave kaon exchange. We compare the contributions of the mass polarization term (the characteristic term of the Schr\"odinger equation) and the exchange term, which is clearly defined in the Faddeev approach in configuration space. The exchange interaction becomes the maximal for a system with two identical particles, such as the $KK{\bar K}$ system. Thus, the bosonic  model for $KK{\bar K}$ system leads to a strongly bound state of this system. A model with a significant isospin splitting of the $K{\bar K}$  potential generates a weakly bounded $KK{\bar K}$ system, because the contribution of the exchange interaction is reduced due to the significant decrease of the $K{\bar K}$ attraction. We have demonstrated that the model for the $KK\bar K$ system with $K\bar K$ interaction having the same strength in the isospin singlet and triplet states is far from Efimov physics. The evaluation of the mass polarization in the $KK{\bar K}$ system in the framework of the Faddeev equations in configuration space allows us to understand, explain and interpret the contribution from the $KK$ potential to the mass of the $K$(1460) as a dynamically generated resonance. It is shown that the contribution of mass polarization into the energy of the $KK{\bar K}$ system is large (up to 30\%) and depends linearly on the $K\bar K$ scattering length. Specifically,  the contribution is defined by the mass ratio of non-identical particles. As a result, relative contributions can be the same for different systems.

We also studied the impact of isospin splitting of the $K\bar K$ interaction on the energy of the $KK\bar K$ quasi-bound state.
Generally, a model with the isospin dependence of a $K\bar K$ potential leads to a decreasing binding energy of the system.
In particular, we found that the $K\bar K$ potential with an essential difference of isospin components 
produces a weak quasi-bound state. The comparison of our calculations with the recent experimental
study 1482.40$\pm $3.58$\pm $15.22 MeV \cite{LHCb}, where the first uncertainty is statistical and the second systematic, shows that the mass of the $K(1460)$ resonance is in a satisfactory agreement with the mass upper bound, calculated within our three-body model with isospin splitting $K{\bar K}$ potential. Due to the experimental uncertainties in the relevant observable, one can explore the possible range for the ratio of the strengths of isospin triplet and singlet components of the $K\bar {K}$ interaction. On the other hand, our results obtained by the model with the same strength of the $K\bar K$ interaction in the isospin singlet and triplet states are in reasonable agreement with the SLAC and ACCMOR collaboration experimental values of the mass of $K$(1460) resonance \cite{Brandenburg,DAUM1981}.

It is worth noticing that despite its simplicity, the single-channel model can reproduce the mass of the $K(1460)$ resonance. In our consideration there are no any fitting parameters, and we are using $s$-wave $K{\bar K}$ and $KK$ two-body potentials and kaons masses only as the inputs in our model. The key ingredient of the model is the proper description of the isospin-dependent $K\bar {K}$ interaction. Therefore, some refinements can be done, such as using more realistic two-body potentials, including $p$-wave components, and/or considering the coupled-channel approach. However, these will not affect dramatically the main conclusions obtained within the present approach.

\acknowledgments
This work is supported by the NSF (HRD-1345219).
\appendix

\section{\label{sec:testing0} Coulomb potential}
The general form of the  Faddeev equations with Coulomb interactions reads as follows \cite{FM}:
\begin{equation}
\{H_0+V^s_{\gamma}(\vert\mathbf{x}_{\gamma}\vert)+
\sum_{\beta =1}^3V^{Coul}_{\beta}(\vert\mathbf{x}_{\beta}\vert)-E\}
\Psi_{\gamma}(\mathbf{x}_{\gamma},\mathbf{y}_{\gamma})
=-V_{\gamma}(\vert\mathbf{x}_{\gamma}\vert)
\sum_{\beta\ne\gamma}\Psi_{\beta}(\mathbf{x}_{\beta},\mathbf{x}_{\beta}),
\label{F0}
\end{equation}
where $V^{Coul}_{\beta}$ is the Coulomb potential between the particles
belonging to the pair $\beta$ and $V_\gamma$ is the short-range
pair potential in the channel $\gamma$, ($\gamma$=1,2,3). In (\ref{F0})
$H_0=-\Delta_{\mathbf{x}_{\gamma}}-\Delta_{\mathbf{y}_{\gamma}}$ is the
kinetic energy operator, $E$ is the total energy, $\Psi$ is the wave
function of the three-body system given as a sum over
three Faddeev components, $\Psi =\sum^3_{\gamma=1}\Psi_\gamma$, $\mathbf{x}_{\gamma}$ and $\mathbf{y}_{\gamma}$ are the Jacobi coordinates
for three particles with unequal masses $m_1$, $m_2$ and $m_3$ having positions $\mathbf{r}_{1},\mathbf{r}_{2}$
and $\mathbf{r}_{3}$ defined as
\begin{align}
\mathbf{x}_{i}& =\sqrt{\frac{m_{j}m_{k}}{m_{j}+m_{k}}}(\mathbf{r}_{j}-%
\mathbf{r}_{k}),\text{ \ }  \notag \\
\mathbf{y}_{i}& =\sqrt{\frac{m_{j}(m_{j}+m_{k})}{M}}\left( -\mathbf{r}_{i}+%
\frac{m_{j}\mathbf{r}_{j}+m_{k}\mathbf{r}_{k}}{m_{j}+m_{k}}\right) ,  \notag
\\
\mathbf{R}& =(m_{1}\mathbf{r}_{1}+m_{2}\mathbf{r}_{2}+m_{3}\mathbf{r}_{3}),%
\text{ }M=m_{1}+m_{2}+m_{3},\text{ }i\neq j\neq k=1,2,3.  \label{3}
\end{align}%
For a system with two identical particles (\ref{F0}) is reduced to two equations.  The system $ K^+K^+K^-$ has two types of Coulomb potentials. The first one is repulsive and describes the interaction between
two particles of the same charge and the second is  attractive and describes the interaction between two opposite charged particles. Each potential gives the contribution to each equation of the set.  For example,
the Coulomb potential of the first type is written as $n_1/|\mathbf{x}|$ for the first equation and $n_2/|\mathbf{x'}|$ for the second equation of the set (\ref{eq:1av}), where $\mathbf{x}'=\mathbf{x}/2+\mathbf{y}$ and $n_k$, $k=1,2$ is reduced charge: $n_k=e^2m_k/\hslash^2$, $m_k$ is a reduced mass of  corresponding particle pair.
\begin{figure}[ht]
\includegraphics[width=15pc]{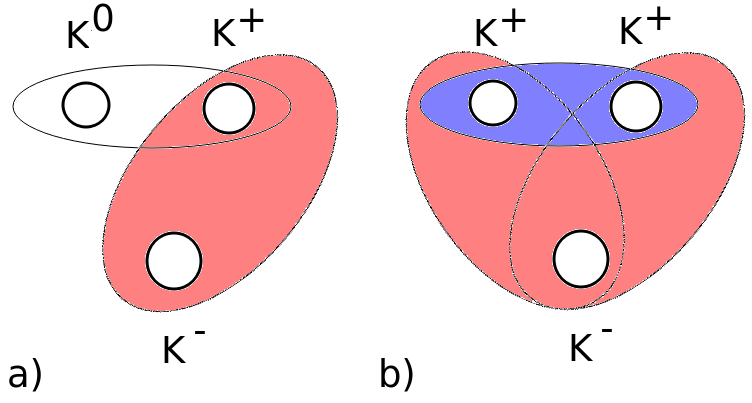}
\caption{\label{fig:C} The structure of the Coulomb force in the particle configurations of the  $KK\bar K$ system, a)  $K^0K^+K^-$, b) $ K^+K^+K^-$.}
\end{figure}

The particle  configurations and corresponding Coulomb forces are schematically presented in Fig. \ref{fig:C}. Note that the Coulomb potential of the particle configuration $K^0K^+K^-$ violates
the paradigm of $AAB$ system. To describe this system with the Coulomb potential, one has to use the Faddeev equations (\ref{F0}) for three non-identical particles.

\section{\label{sec:width} The width of the $KK\bar K$ system}
For  evaluation of the width of the $KK\bar K$ quasi-resonance, we taken into account that the real part of the complex $K\bar K$ potentials dominates with the ratio for strengths of $\epsilon=283/1155$ for the potential of the A parameter set and $\epsilon=210/630$ for the potential with set B. One can write the Hamiltonian of the $KK\bar K$ system as
$H^R+i\epsilon \text{Im} V_{K\bar K}$, where  $H^R=H_0+\text{Re} V_{K\bar K}+V_{KK}$
We have taken into account that
the $KK$ potential has no an imaginary part.
This complex-value expression for the Hamiltonian can be transformed into the real 2$\times$2 matrix representation:
$$
  \left( {\begin{array}{cc}
 H^R    &      -\epsilon \text{Im} V_{K\bar K} \\
          \epsilon \text{Im} V_{K\bar K}       &  H^R\\
  \end{array} } \right).
$$
    The  obtained matrix is a rotation-scaling matrix. The complex eigenvalues of the matrix are $H^R\pm i\epsilon \text{Im} V_{K\bar K}$.
The energy $E$ and width $\Gamma$ can be obtained by the averaging $E=\langle  H^R \rangle \pm i \epsilon \langle V_{K\bar K} \rangle =E^R\pm i\Gamma/2$.
We have  evaluated the averaged  $K{\bar K}$ potential energy as $\langle V_{K\bar K} \rangle$ for the $K^0K^0{\bar K}^0$ particle configuration. The corresponding results using potentials with the A and B parameter sets are presented in Table
\ref{tab-3a}. A similar result for the parameter set A was obtained in Ref. \cite{MartJidoKanada}, where the variational calculations have been performed.

\section{\label{sec:testing} Numerical solution of the Faddeev equations: the code testing }
\subsection{\label{sec:testing1} Bosonic model for $nnp$ system with MT-V potential}
\begin{figure}[ht]
\includegraphics[width=19 pc]{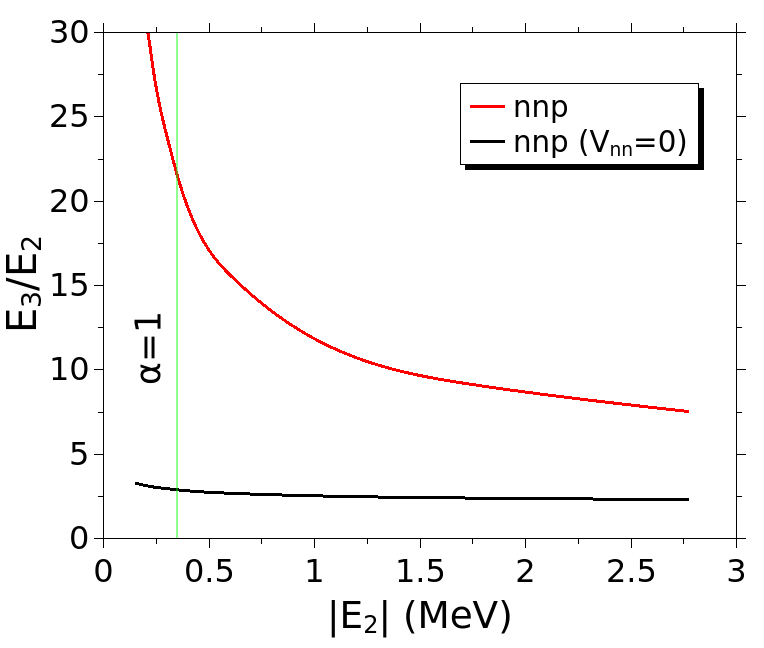}
\includegraphics[width=19 pc]{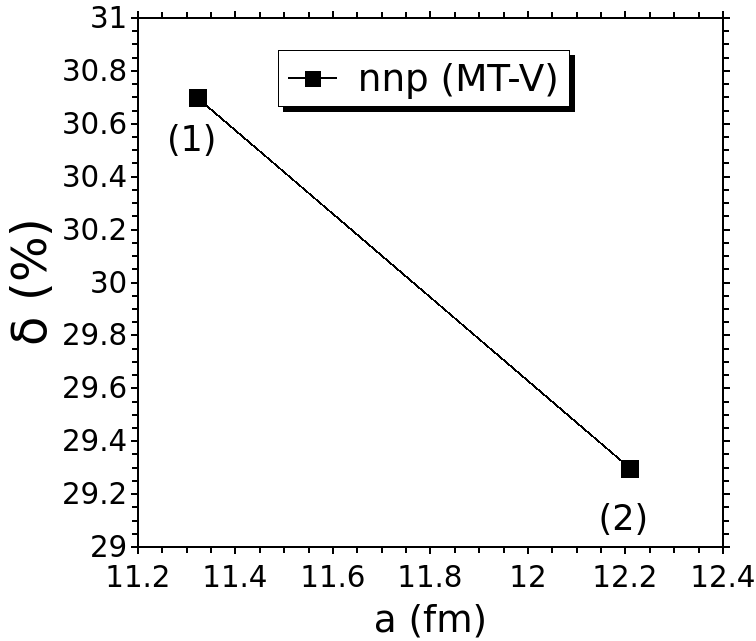}
\caption{\label{fig:Ed} The ratio $E_3/E_2$ with dependence on two-body energy $|E_2|$  for the $nnp$ system calculated with the MT-V (1) potential.
The parameter is factor $\alpha$ defining the scaled $nn$ potential $V_{nn}\to \alpha V_{nn}$. The vertical line corresponds to $\alpha=1$. (Left panel). The correlations between the relative contribution of the mass polarization $\delta$ and  $np$-scattering length $a$ in $nnp$ systems calculated with two versions of $NN$ MT-V potential. (Right panel).
}
\end{figure}

Our calculation for $KK\bar K$ system is tested by using the simple model for $nnp$ system with the MT-V nucleon-nucleon potential \cite{F2}. The potential corresponds to
a bosonic model for $nnp$ system, when an isospin/spin independent $s$-wave potential is used. The MT-V bosonic model was motivated by spin averaging for the spin-dependent MT-III potential \cite{MT}.
The configuration space  Faddeev calculations for the model are based on Eqs. (\ref{eq:1av}). The
MT-V potential is defined as
$V_{NN}(r)=\sum\limits_{i=1,2}U_{i}\exp \left[ -\left( \mu_i r\right) \right]/r$, where $U_{i}$ and $\mu$ are  strength  and  range parameters, respectively.
The range parameters are $\mu_1=1.55$ fm$^{-1}$, $\mu_1=3.11$ fm$^{-1}$.
We used two sets for strength parameters of the potential  known from literature \cite{F}: (1) $U_{1}=-570.316$ MeV, $U_{2}=1438.4812$ MeV and
(2)  $U_{1}=-578.098$ MeV, $U_{2}=1458.047$ MeV.
The results of the calculations are given in Table. \ref{tab-4}. Our results are in good agreement with the results obtained in Ref. \cite{F}, where the Faddeev equations in configuration space were also applied.
Note, that  the mass polarization evaluated by the value $\Delta/|E_3|(V_{KK}=0)$ is similar to the kaonic system  $KK\bar K$ due to the similar mass ratio.
\begin{table}[ht]
\centering
\caption{\label{tab-4} The energy ($E_3$) of bound state of $nnp$ system  within different variants for MT-V nucleon-nucleon potential.   The $E_2$ is  energy of bound $np$ pair.
 The  energies are given in  MeV.  The relative contribution of the mass polarization (MP) $\delta=\Delta/|E_3(V_{KK}=0)|$ is shown.
 The results of calculations from \cite{F} are listed in  parentheses. }
\begin{tabular}{@{}*{6}{ccccc}}
 \hline
Potential &       $E_3$  &       $E_3(V_{nn}=0)$  &  $ E_2$       & MP  (\%)\\  \hline
MT-V(1) &   -7.54  (-7.54)      &         -1.01                            &         -0.35            &  30.7 \\
MT-V(2) &    -8.04  (-8.0424)     &        -1.16                             &   -0.41              &   29.3\\
 \hline
\end{tabular}
\end{table}
For the bosonic  model  of the $nnp$ system, the mentioned above (Section \ref{sec:results}) correlations between two and three-body parameters takes place. In particular,  the ratio $E_3/E_2$ with dependence on two-body energy $|E_2|$  presented in Fig. \ref{fig:Ed} shows the Efimov effect when two-body energy is close to three-body threshold.
Here,  the $nn$ potential is scaled by a factor $\alpha$ as $V_{nn}\to \alpha V_{nn}$ with the condition  $\alpha>$0. The strong attraction of the $nn$ pair makes the ratio  $E_3/E_2$ larger in comparison  with the  $E_3(V_{nn}=0)/E_2$ case, when the $nn$ interaction is omitted.
The results of our calculations for the $nnp$ system with two sets of the  MT-V potential for the correlation between the relative contributions of the mass polarization $\delta$ and
two-body scattering length $a$ are is are represented in Fig. \ref{fig:Ed}. A similar dependence is shown in Fig. \ref{fig:0}  obtained
in the case of the $KK\bar K$ and $\alpha \Lambda \Lambda$ systems.
The correlation between of $\delta$ and $a$ is represented by a linear dependence with the negative slope as one can see in Fig. \ref{fig:0}.

\subsection{Cluster reduction method versus direct numerical solution}
  The cluster reduction method \cite{CRM,CRM1} was alternatively used for a numerical solution of the Faddeev equations  (\ref{eq:1comp}). The method is based on the expansion of the components $U$ and $W$ in terms of the basis of the eigenfunctions of two-body Hamiltonian of the subsystems:
\begin{equation}
U(x,y) = \sum_{i\ge 1}^N \phi^{U}_{i}(x)F^{U}_{i}(y), \quad
W(x,y) = \sum_{i\ge 1}^N \phi^{W}_{i}(x)F^{W}_{i}(y).
\label{Ex0}
\end{equation}
Here, the functions  $F^U_{i}$ and $F^W_{i}$, $i=1,2,\dots$, $N$ describe  the relative motion of ''clusters'' in each rearrangement channel
$(KK)\bar K$ and $K(K\bar K)$, respectively. The  functions  $F^U_{i}$ ($F^W_{i}$) depend  on the
relative  coordinate  $y$.
The  solutions of the two-body Schr\"odinger equations form a complete set of eigenfunctions in the box,
$x \subset [0,R_x]$:
$$
(-\frac{\hbar ^{2}}{2\mu^U}\partial _{x}^{2}+V^{s_{NN}=0}_{NN}(x))\phi^U_i(x)~=~\epsilon^U_i\phi^U_i(x), \qquad
(-\frac{\hbar ^{2}}{2\mu^W}\partial _{x}^{2}+V_{N{\bar K}}(x))\phi^W_i(x)~=~\epsilon^W_i\phi^W_i(x),
$$
where, $\mu^U$ (and $\mu^W$) is a reduced mass of the pairs and
$\phi^U_i(0)=\phi^U_i(R_x)=0$ ($\phi^W_i(0)=\phi^W_i(R_x)=0$), $\quad i=1,2,\dots, N$. The parameter $R_x$ is chosen to be large enough to reproduce the pair binding energy. In our calculations $R_x$=35~fm is used. The number $N$ is chosen by condition of total convergence of calculations results, when $N$ consequently increases.

\begin{figure}[ht]
\includegraphics[width=25pc]{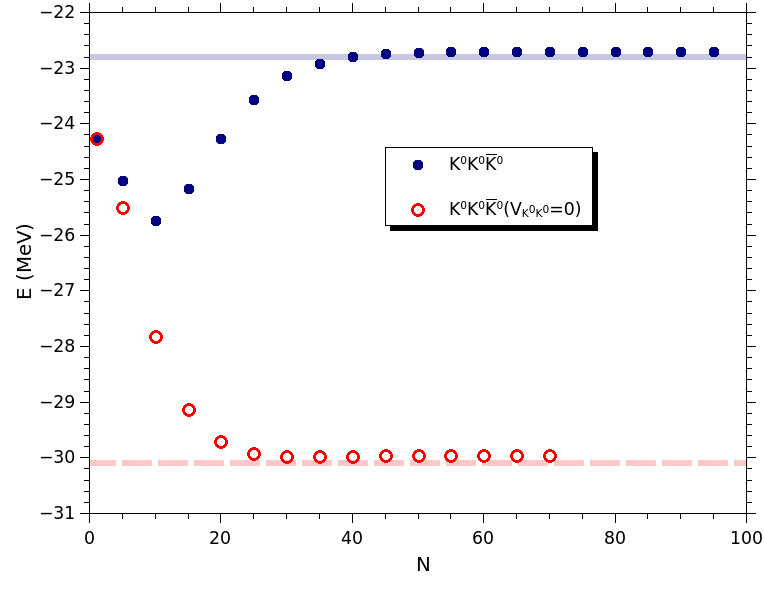}
\caption{ \label{fig:EN} The binding energy of the $KK\bar K$ system ($K^0K^0\bar{K^0}$) calculated using CRM for different numbers of the terms in the expansion (\ref{Ex0}). The case when the  $KK$ potential is omitted is also shown. The horizontal lines (solid and dashed) represent the results of the direct numerical solution of the Faddeev equations. The parameter set B of the potentials was applied. }
\end{figure}

The  comparison of the CRM and direct solution is presented in Fig. \ref{fig:EN}. The results obtained using the both methods are in good agreement.
The CRM calculations for the case of complete set of potentials and the case of restriction $V_{K^0K^0}=0$ demonstrate the repulsive nature of the  $KK$ potential. The convergence of the calculation results, as a function of the number $N$ of the terms in Eq. (\ref{Ex0}) is different for these cases. In the first case, the decrease of binding energy changes to increase when the calculation becomes "more precise", by increasing the number $N$. For the second case, we have a monotonic decrease in the binding energy.  Such behavior is related to the consequent inclusion of the attraction for the $K\bar K$ pair  and repulsion for the  $KK$ pair.


\begin{thebibliography}{99}
\bibitem{AliProg2017} A. Ali, J. S. Lange, and S. Stone, Exotics: Heavy pentaquarks and tetraquarks. Prog. Part. Nucl.
Phys. \textbf{97}, 123 (2017).

\bibitem{Chen2017} H.-X. Chen, W. Chen, X. Liu, Y.-R. Liu, and S.-L. Zhu, A review of the open charm and open bottom systems.
Rep. Prog. Phys. \textbf{80}, 076201 (2017).

\bibitem{Olsen} S. L. Olsen, T. Skwarnicki, and D. Zieminska, Nonstandard heavy mesons and baryons: Experimental evidence. Rev. Mod.
Phys. \textbf{90}, 015003 (2018).

\bibitem{Mei2018} F.-K. Guo, C. Hanhart, U-G. Mei\ss ner, Q. Wang, Q. Zhao,
and B.-S. Zou, Hadronic molecules. Rev. Mod. Phys. \textbf{90}, 015004 (2018).

\bibitem{Nielsen} M. Nielsen, F. S. Navarra, and S. H. Lee, New charmonium
states in QCD sum rules: a concise review, Phys. Rep. \textbf{497}, 41
(2010).

\bibitem{Nakahara} Y. Nakahara, M. Asakawa, and T. Hatsuda, Hadronic
spectral functions in lattice QCD, Phys. Rev. D \textbf{60}, 091503 (1999).

\bibitem{Bazavov} A. Bazavov, et al. Nonperturbative QCD simulations with
2+1 flavors of improved staggered quarks. Rev. Mod. Phys. \textbf{82}. 1349--1417 (2010).

\bibitem{Gasser} J. Gasser and H. Leutwyler, Chiral perturbation theory:
expansions in the mass of the strange quark, Nucl. Phys. B \textbf{250}, 465
(1985).

\bibitem{2} U-G. Mei\ss ner, Recent developments in chiral perturbation theory. Rep. Prog. Phys. \textbf{56}, 903 (1993).

\bibitem{3} A. Pich, Chiral perturbation theory. Rep. Prog. Phys. \textbf{58}, 563 (1995).

\bibitem{1} J. A. Oller, E. Oset, and J. R. Pelaez, Nonperturbative approach to effective chiral Lagrangians and meson interactions. Phys. Rev. Lett. \textbf{%
80}, 3452 (1998).

\bibitem{Dosch} H. G. Dosch, Gluon condensate and effective linear potential. Phys. Lett. B \textbf{190}, 177 (1987).

\bibitem{Simonov} H. G. Dosch and Yu. A. Simonov, The area law of the Wilson loop and vacuum field correlators. Phys. Lett. B \textbf{205}, 339 (1988).

\bibitem{Simonov2} Yu. A. Simonov, Vacuum background fields in QCD as a source of confinement. Nucl. Phys. B \textbf{307}, 512 (1988);

\bibitem{11} A. Di Giacomo, H. G. Dosch, V. I. Shevchenko, and Yu. A.
Simonov, Field correlators in QCD. Theory and applications. Phys. Rep. \textbf{372}, 319 (2002).


\bibitem{Bashir} A. Bashir, L. Chang, I. C. Cloet, B. El-Bennich,Y. Liu, C. D. Roberts, and P. C. Tandy, Collective perspective on advances in Dyson-Schwinger equation QCD. Commun. Theor. Phys. \textbf{58}, 79 (2012).

\bibitem{Roberts1}C. D. Roberts and A. G. Williams, Dyson-Schwinger equations and the application to hadronic physics. Prog. Part. Nucl. Phys. \textbf{33}, 477 (1994).

\bibitem{Roberts2} C. D. Roberts and S. M. Schmidt, Dyson-Schwinger equations: Density, temperature and continuum strong QCD. Prog. Part. Nucl. Phys. \textbf{45}, S1 (2000).

\bibitem{Roberts3} C. D. Roberts, Hadron properties and Dyson-Schwinger equations. Prog. Part. Nucl. Phys. \textbf{61}, 50 (2008).

\bibitem{Roberts4} R. J. Holt and C. D. Roberts, Nucleon and pion distribution functions in the valence region. Rev. Mod. Phys. \textbf{82}, 2991 (2010).

\bibitem{Cyrol1} A. K. Cyrol, M. Mitter, J. M. Pawlowski, and N. Strodthoff, Nonperturbative quark, gluon, and meson correlators of unquenched QCD. Phys. Rev. D \textbf{97}, 054006 (2018).

\bibitem{Cyrol2} M. Mitter, J.M. Pawlowski, and N. Strodthoff, Chiral symmetry breaking in continuum QCD. Phys. Rev. D \textbf{91}, 054035 (2015).

\bibitem{Cyrol3} A. K. Cyrol, L. Fister, M. Mitter, J. M. Pawlowski, and N. Strodthoff, Landau gauge Yang-Mills correlation functions. Phys. Rev. D \textbf{94}, 054005 (2016).

\bibitem{Eichmann} G. Eichmann, H. Sanchis-Alepuz, R. Williams, R. Alkofer, C. S. Fischer, Baryons as relativistic three-quark bound states. Prog. Part. Nucl. Phys. \textbf{91}, 1 (2016).

\bibitem{Hiller} J.R. Hiller, Nonperturbative light-front Hamiltonian methods. Prog. Part. Nucl. Phys. \textbf{90}, 75 (2016).

\bibitem{Fad} L.D. Faddeev, Scattering theory for a three-particle system. ZhETF \textbf{39}, 1459 (1961); [Sov. Phys. JETP \textbf{12}, 1014 (1961)].

\bibitem{Fad1}L.D. Faddeev, Mathematical problems of the quantum theory of
scattering for a system of three particles. Proc. Math. Inst. Acad. Sciences USSR \textbf{69}, 1-122 (1963).

\bibitem{Delves2} L. M. Delves, Tertiary and general-order collisions (II), Nucl. Phys. \textbf{20}, 275-308 (1960).

\bibitem{Smith} F. T. Smith, Generalized Angular Momentum in Many-Body Collisions, Phys. Rev. \textbf{120}, 1058-1069 (1960).

\bibitem{Delves3} L. M. Delves, Three-particle photo-disintegration of the triton, Nucl. Phys. \textbf{29}, 268-280 (1962).

\bibitem{SimonovHH} Yu. A. Simonov, The three body problem. A complete system of angular functions. Yad. Fiz. \textbf{3}, 630-638 (1966); [Sov. J. Nucl. Phys. \textbf{3}, 461-466 (1966)].

\bibitem{SimonovHH2} A. M. Badalyan and Yu. A. Simonov, The three body problem. Equation for the partial waves, Yad. Fiz. \textbf{3}, 1032 (1966); [Sov.J.Nucl.Phys. \textbf{3}, 755-764 (1966)]

\bibitem{JKez1980} R. I. Jibuti and R. Ya. Kezerashvili, On the theory of "true" three particle scattering. Czech. J. Phys., B \textbf{30}, 1090-1100 (1980).

\bibitem{Kez1983} R. Ya. Kezerashvili, Theory three-body to three-body elastic scattering in
the hyperspherical formalism. Yad. Fiz. \textbf{3}, 1032 (1966);[Sov.J. Nucl. Phys. \textbf{38}, 293-298 (1983)].

\bibitem{JKez1985} R. I. Jibuti and R. Ya. Kezerashvili, Double-charge-exchange of $\pi $%
-mesons on three- and four-particle nuclei. Nucl. Phys.  A \textbf{437}
687-716 (1985).

\bibitem{Roskez1993} R. Ya. Kezerashvili and S. Rosati, On the asymptotic behavior of the elastic
3$\rightarrow $3 scattering wave function in the hyperspherical basis, Phys.
Lett. B \textbf{318}, 23-25 (1993).

\bibitem{Kez1994} R. Ya. Kezerashvili, On the asymptotic behaviour of the 4 4 scattering wave
function in the hyperspherical representation, Phys. Lett. B \textbf{334},
263-267 (1994).

\bibitem{Kez2001} R. Ya. Kezerashvili, Elastic $N$-body to $N$-body scattering in hyperspherical
representation. World Scientific. Adv. in Quantum Many-Body Theory. \textbf{5}, 89-95 (2001).

\bibitem{Koslovsky2012} V. I. Kovalchuk and I. V. Kozlovsky, Faddeev equations and the method of hyperspherical
harmonics in the problem of three-nucleon continuum. Phys. Part. Nucl. |textbf{43}, 294-310 (2012); [Fiz. Element. Chast. i Atom. Yad. \textbf{43}, 573-610 (2012)].

\bibitem{Kievsky2018} J. Dohet-Eraly, M. Viviani, A. Kievsky, L. E. Marcucci, and A. Gnech,
Few-Nucleon Systems with an Hyperspherical Harmonic Method, Recent Prog.
Few-Body Phys. \textbf{238}, 163-167 (2018).

\bibitem{Jkrup1977} R. I. Jibuti , N. B. Krupennikova, and V. Yu. Tomchinsky, Hyperspherical
basis for the continuum spectrum. Nucl. Phys. A \textbf{276}, 421-432 (1977).

\bibitem{AGS}E. O. Alt, P. Grassberger, and W. Sandhas,
Reduction of the three - particle collision problem to multichannel two - particle Lippmann-Schwinger equations. Nucl. Phys. B \textbf{2}, 167 (1967).

\bibitem{4} A. Martinez Torres, K. P. Khemchandani, and E. Oset, Three-body resonances in two-meson one-baryon systems. Phys. Rev.
C \textbf{77}, 042203(R) (2008).

\bibitem{41} A. Martinez Torres, K. P. Khemchandani, L. S. Geng, M.
Napsuciale and E. Oset, $X(2175)$ as a resonant state of the $\phi K\bar K$
 system. Phys. Rev. D \textbf{78}, 074031 (2008).

\bibitem{17} A. Martinez Torres, K. P. Khemchandani, and E. Oset, Solution to Faddeev equations with two-body experimental amplitudes as input and application to $J^{P}=1/2^{+}$, $S=0$ baryon resonances. Phys. Rev.
C \textbf{79}, 065207 (2009).

\bibitem{18} A. Martinez Torres and D. Jido, $K\lambda (1405)$ configuration of the $K\bar KN$ system. Phys. Rev. C \textbf{82},
038202 (2010).

\bibitem{5} L. Roca and E. Oset, Description of the $f_{2}(1270)$, $\rho_{3}(1690)$, $f_{4}(2050)$, $\rho_{5}(2350)$, and $f_{6}(2510)$ resonances as multi-$\rho(770)$ states. Phys. Rev. D \textbf{82}, 054013 (2010).

\bibitem{6} L. Roca, Pseudotensor mesons as three-body resonances. Phys. Rev. D \textbf{84}, 094006 (2011).

\bibitem{7} A. Martinez Torres, E. J. Garzon, E. Oset, and L. R. Dai, Limits to the fixed center approximation to Faddeev equations: The case of the
$\phi(2170)$. Phys.
Rev. D \textbf{83}, 116002 (2011).

\bibitem{Bayar} M. Bayar, J. Yamagata-Sekihara and E. Oset, $\bar KNN$  system with chiral dynamics. Phys. Rev. C
\textbf{84}, 015209 (2011).

\bibitem{Bayar2} M. Bayar, C. W. Xiao, T. Hyodo, A. Dote, M. Oka and E.
Oset, Energy and width of a narrow $I=1/2$
$DNN$  quasibound state. Phys. Rev. C \textbf{86} (2012) 044004

\bibitem{8} C. W. Xiao, M. Bayar, and E. Oset, Prediction of $D^{*}$-multi-$\rho$ states. Phys. Rev. D \textbf{86},
094019 (2012).

\bibitem{9} B. Durkaya and M. Bayar, Faddeev fixed-center approximation to the $\rho\bar DD$ system. Phys. Rev. D \textbf{92}, 036006 (2015).

\bibitem{91} M. Bayar, X. L. Ren and E. Oset, States of $B^{*}\bar B ^{*}$ with $J = 3$ within the fixed center approximation to Faddeev equations. Eur. Phys. J. A \textbf{51},
61 (2015).

\bibitem{92} M. Bayar, X. L. Ren, B. Durkaya, and E. Oset, The X(4260) and
X(4360) mesons as a $\rho D\overline{D}$ molecular state and states of $\rho
D^{\ast }\overline{D}^{\ast }$with $J=3$. AIP Conf. Proc. \textbf{1735},
050009 (2016).

\bibitem{922} J. M. Dias, V. R. Debastiani, L. Roca, S. Sakai, and E. Oset, Binding of the $BD\bar D$ and $BDD$ systems.
Phys. Rev. D \textbf{9}6, 094007 (2017).

\bibitem{93} X. L. Ren and Z. Sun, Possible bound states with hidden
bottom from $\bar K^{(\ast)}B^{(\ast )}\bar B^{(\ast )}$systems.
Phys. Rev. D \textbf{99}, 094041 (2019).

\bibitem{12} A. Martinez Torres, K. P. Khemchandania, D. Jido, Y.
Kanada-Eniyo, and E. Oset, Three-body hadron systems with strangeness. Nucl. Phys. A \textbf{914}, 280-288 (2013).

\bibitem{13} E. Oset, A. Martinez Torres, K.P. Khemchandani, L. Roca, J.
Yamagata, Two, three, many body systems involving mesons.
Progress in Part. and Nucl. Phys. \textbf{67}, 455 (2012).

\bibitem{101} I. M. Narodetskii and M. A. Trusov, Ground-state baryons in nonperturbative quark dynamics. Yad. Fiz. \textbf{67}, 783
(2004) [Phys. At. Nucl. \textbf{67}, 762 (2004)].

\bibitem{102} O. N. Driga, I. M. Narodetskii, and A. I. Veselov, Pentaquarks in the Jaffe-Wilczek approximation. Yad. Fiz.
\textbf{71}, 356 (2008) [Phys. At. Nucl. \textbf{71}, 335 (2008)].

\bibitem{103} I. M. Narodetskii, C. Semay, and A. I. Veselov, Accuracy of auxiliary field approach for baryons. Eur. Phys. J.
C \textbf{55}, 403 (2008).

\bibitem{10} R. Ya. Kezerashvili, I. M. Narodetskii, and A. I. Veselov, Baryons in the field correlator method: effects of the running strong coupling.
Phys. Rev. D \textbf{79}, 034003 (2009).

\bibitem{10P} R. Ya. Kezerashvili, I. M. Narodetskii, and A. I. Veselov, Baryons in the Field Correlator Method, AIP Conf. Proc. \textbf{1182}, 526 (2009).

\bibitem{Brandenburg} G. W. Brandenburg, et al., Evidence for a new strangeness-one pseudoscalar meson. Phys. Rev. Lett. \textbf{36}, 1239 (1976).


\bibitem{DAUM1981} C. Daum, et al., Diffractive production of strange mesons at 63 GeV. Nucl. Phys. B \textbf{187}, 1-4 (1981).

\bibitem{PDG2018} M. Tanabashi et al., Particle Data Group 2018, Reviewe of particle physics. Phys. Rev. D
\textbf{98,} 030001 (2018).

\bibitem{LHCb} LHCb collaboration, R. Aaij et al., Studies of the resonance structure in $D^{0}\rightarrow K^{\mp }\pi ^{\pm }\pi ^{\pm }\pi ^{\pm }$ decays. Eur. Phys. J. C \textbf{78%
}, 443 (2018).

\bibitem{Izgur85} S. Godfrey and N. Isgur, Mesons in a relativized quark model with chromodynamics. Phys. Rev. D \textbf{32}, 189
(1985).

\bibitem{MartJidoKanada} A. Martinez Torres, D. Jido, and Y. Kanada-En'yo, Theoretical study of the
$KK\bar K$  system and dynamical generation of the $K(1460)$ resonance.
Phys. Rev. C \textbf{83}, 065205 (2011).

\bibitem{KanadaJido} Y. Kanada En'yo, D. Jido, $\bar K\bar KN$ molecular state in a three-body calculation. Phys. Rev. C \textbf{78},
025212 (2008).

\bibitem{JidoKanada} D. Jido, Y. Kanada-En'yo, $\bar KKN$ molecule state with $I=1/2$
 and $J^{P}=1/2^{+}$  studied with a three-body calculation. Phys. Rev C \textbf{78},
035203 (2008).

\bibitem{Albaladejo} M. Albaladejo, J. A. Oller, and L. Roca, Dynamical generation of pseudoscalar resonances. Phys. Rev. D
\textbf{82}, 094019 (2010).

\bibitem{RKSh.Ts2014} R. Ya. Kezerashvili and Sh. M. Tsiklauri, Study of the $KK\bar K$ system in hyperspherical formalism. EPJ Web
Conf. \textbf{81}, 02022 (2014).

\bibitem{SHY19} S. Shinmura, K. Hara and T. Yamada, Effects of attractive $K\bar K$ and repulsive $KK$ interactions in $KK\bar K$ three-body resonance. Proc. 8th Int. Conf.
Quarks and Nuclear Physics (QNP2018), JPS Conf. Proc., 023003 (2019).

\bibitem{Fed1}E. Nielsen, D. V. Fedorov, A. S. Jensen, and E. Garrido, The three-body problem with short-range interactions. Phys. Rep. \textbf{347}, 373-459 (2001).

\bibitem{Fed2}A. S. Jensen, K. Riisager, D. V. Fedorov, and E. Garrido, Structure and reactions of quantum halos. Rev. Mod. Phys. \textbf{76}, 215-261 (2004).

\bibitem{FM} L.D. Faddeev and S.P. Merkuriev, {\ Quantum Scattering Theory
for Several Particle Systems} (Kluwer Academic, Dordrecht, 1993) pp. 398.

\bibitem{K1991} Yu. A. Kuperin, D. M. Latypov, S. P. Merkuriev, M. Bruno, and F.
Cannata,  $d\alpha$ scattering in three body model. Sov. J. Nucl. Phys. \textbf{53}, 582 (1991).


\bibitem{K2015} R.Ya. Kezerashvili, S.M. Tsiklauri, I. Filikhin, V.M. Suslov
and B. Vlahovic, Three-body calculations for the $K^{-}pp$ system within potential models. J. Phys. G: Nucl. Part. Phys. \textbf{43}, 065104 (2016).

\bibitem{Revai} J. Revai, Three-body calculation of the 1s level shift in kaonic deuterium
with realistic $KN$ potentials, Phys. Rev. C \textbf{94}, 054001 (2016).

\bibitem{Igor2019}I. Filikhin, V.M. Suslov, B. Vlahovic, Nd breakup within isospinless AAB model, J. Phys. G: Nucl. Part. Phys. \textbf{46}, 105103 (2019)

\bibitem {AYPRC2002}Y. Akaishi and T. Yamazaki, Nuclear $\bar K$ bound states in light nuclei. Phys. Rev. C65, 044005 (2002).

\bibitem {AYKN}T. Yamazaki and Y. Akaishi, Basic $\bar K$ nuclear cluster, $K^{-}pp$, and its enhanced formation in the $p+p \rightarrow K^{+}+X$ reaction. Phys. Rev. C \textbf{76}, 045201 (2007).

\bibitem {DHW}A. Dot\'{e}, T. Hyodo and W. Weise, Variational calculation of the $ppK^{-}$ system based on chiral SU(3) dynamics. Phys. Rev. C \textbf{79},
014003 (2009).

\bibitem {QCDcalculKK}S. R. Beane, T. C. Luu, K. Orginos, A. Parreno, M. J.
Savage, and A. Torok, $K^{+}K^{+}$ scattering length from lattice QCD. Phys. Rev. D \textbf{77}, 094507 (2008).

\bibitem{H2002} E. Hiyama, M. Kamimura, T. Motoba, T. Yamada and Y.
Yamamoto, Four-body cluster structure of A=7-10 double-$\Lambda$ hypernuclei. Phys. Rev. C \textbf{66}, 024007-13 (2002).

\bibitem{FG2002} I.N. Filikhin and A. Gal, Light $\Lambda\Lambda$ hypernuclei and the onset of stability for $\Lambda\Xi$ hypernuclei. Phys. Rev. C \textbf{65},
041001(R)-4 (2002).

\bibitem{F2018} I. Filikhin, R.Ya. Kezerashvili, V.M. Suslov and B.
Vlahovic, On mass polarization effect in three-body systems,
Few-Body Syst. \textbf{59}, 33 (2018).

\bibitem{NF1968} H. P. Noyes and H. Fiedeldey, in Three-Particle Scattering in Quantum Mechanics (Proc. Texas A$\&$M Conf.),
Ed. by J. Gillespie and J. Nuttall (Benjamin, New York, 1968), p. 195.

\bibitem{MGL} S. P. Merkuriev, C. Gignoux, and A. Laverne, Three-body scattering in configuration space. Ann. Phys. \textbf{99}, 30 (1976).

\bibitem{Su1996} J. Bernabeu, V. M. Suslov, T. A. Strizh, et al. New approach for numerical solution of configuration-space Faddeev equations. Hyperfine Interact \textbf{101}, 391 (1996).

\bibitem{Sus20} V. M. Suslov, I. Filikhin, B. Vlahovic, and M. A. Braun, $nd-$scattering within MGL approach
for configuration-space Faddeev equations. Phys. Part. Nucl. \textbf{50}, 433 (2019).

\bibitem{CRM} S.L. Yakovlev and I.N. Filikhin,
Cluster Reduction of the four-body Yakubovsky equations in configuration space for the bound-state problem and for low-energy scattering,
Phys. At. Nucl. \textbf{60}, 1794 (1997).

\bibitem{CRM1} I.N. Filikhin and S.L. Yakovlev, Calculation of the binding energy and of the parameters of low-energy scattering in the $\Lambda np$ system. Phys. At. Nucl. \textbf{63}, 223 (2000).

\bibitem{Ham2010} H.-W. Hammer and L. Platter, Efimov states in nuclear and particle physics. Annu. Rev. Nucl. Part. Sci. \textbf{60}, 2007 (2010).

\bibitem{NE17} P. Naidon, and S. Endo, Efimov physics: a review. Rep. Prog. Phys. \textbf{80}, 056001
(2017).

\bibitem{PP10} L. Pricoupenko and P. Pedri, Universal $(1+2)-$body bound states in planar atomic waveguides. Phys. Rev. A \textbf{82},  033625 (2010).

\bibitem{BGelman} T. D. Cohen, B. A. Gelman, S. Nussinov, New near-threshold mesons. Phys. Lett. B \textbf{578} 359 (2004).

\bibitem{Ham2008} H.-W. Hammer, Universality in QCD and halo nuclei. Proc. Science, \textbf{8}, 147 (2008).

\bibitem{FV} I. Filikhin and B. Vlahovic, Lower bound for $ppK^{-}$ quasi-bound state energy.  Phys. Part. Nucl. \textbf{51}, 979-987 (2020).

\bibitem{F} J. L. Friar, B.F. Gibson, and G.L. Payne, Configuration space Faddeev calculations. V. Variational bounds. Phys. Rev. C \textbf{24}
, 2279 (1981).


\bibitem{MT} R. A. Malfliet and J. A. Tjon, Solution of the Faddeev equations for the trion problem using local two-particle interactions. Nucl. Phys. A \textbf{127}, 161 (1969).

\bibitem{F2} J.L. Friar, B.F. Gibson, G. Berthold, et al.,
Benchmark solutions for a model three-nucleon scattering problem. Phys. Rev. C \textbf{42}, 1838 (1990).


\end{thebibliography}
\end{document}